\documentclass[runningheads]{llncs}

\usepackage{eccv}

\usepackage{eccvabbrv}

\usepackage{graphicx}
\usepackage{wrapfig}
\usepackage{multirow}
\usepackage{algorithm}
\usepackage{amsmath}
\usepackage{amsfonts}
\usepackage{algpseudocode}
\usepackage{floatflt}
\usepackage{wrapfig}
\usepackage{bbding}
\usepackage{booktabs}
\usepackage{amsmath}
\usepackage{amssymb}
\usepackage{tabularx}
\usepackage[accsupp]{axessibility}  

\usepackage{hyperref}

\usepackage{orcidlink}

\begin{document}

\title{A Diffusion Model for Simulation Ready Coronary Anatomy with Morpho-skeletal Control}

\titlerunning{Morpho-skeletal Diffusion Models}


\author{Karim Kadry\inst{1}\orcidlink{0000-0001-9520-0504} \and
Shreya Gupta\inst{1} \and
Jonas Sogbadji\inst{1} \and
Michiel Schaap \inst{3} \orcidlink{0000-0002-4640-7805} \and
Kersten Petersen \inst{3} \orcidlink{0000-0003-4887-6477} \and 
Takuya Mizukami \inst{4}  \orcidlink{0000-0003-1086-8746} \and 
Carlos Collet \orcidlink{0000-0003-0227-0082} \inst{5} \and
Farhad R. Nezami\inst{2}\orcidlink{0000-0002-4210-3177} \and
Elazer R. Edelman\inst{1,2}\orcidlink{0000-0002-7832-7156}}
\authorrunning{K.~Kadry et al.}


\institute{
Institute of Medical Engineering, MIT \and Brigham and Women's Hospital, Harvard \and HeartFlow \and Showa University \and OLV Aalst \\
\email{kkadry@mit.edu; shreyag@mit.edu; jonassog@mit.edu; mschaap@heartflow.com; kpetersen@heartflow.com; mizukami.tky@gmail.com; carloscollet@gmail.com; frikhtegarnezami@bwh.harvard.edu; ere@mit.edu; }
}
\maketitle

\begin{abstract}
Virtual interventions enable the physics-based simulation of device deployment within coronary arteries. This framework allows for counterfactual reasoning by deploying the same device in different arterial anatomies. However, current methods to create such counterfactual arteries face a trade-off between controllability and realism. In this study, we investigate how Latent Diffusion Models (LDMs) can custom synthesize coronary anatomy for virtual intervention studies based on mid-level anatomic constraints such as topological validity, local morphological shape, and global skeletal structure. We also extend diffusion model guidance strategies to the context of morpho-skeletal conditioning and propose a novel guidance method for continuous attributes that adaptively updates the negative guiding condition throughout sampling. Our framework enables the generation and editing of coronary anatomy in a controllable manner, allowing device designers to derive mechanistic insights regarding anatomic variation and simulated device deployment.
\end{abstract}

\section{Introduction}
\label{sec:intro}

\begin{figure}
  \begin{center}
    \includegraphics[width=1\textwidth]{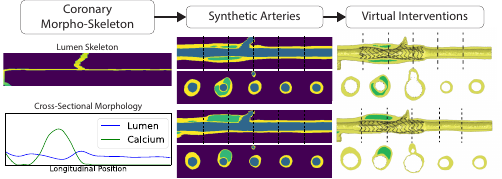}
  \end{center}
  \caption{\small We propose to control the generation of 3D multi-component diseased coronary arteries with mid-level representations such as cross-sectional morphology and tree-like skeletal structure. Synthetic arteries enable physics-based counterfactual reasoning through comparative virtual intervention studies. We synthesize semantic segmentation maps composed of luminal (blue), arterial wall (yellow) and calcified (green) tissues.}
  \label{fig:teaser}
\end{figure}

Coronary artery disease is caused by the buildup of diseased tissue in the arteries that supply blood to the heart \cite{ralapanawa2021epidemiologycoronary}.  Variation in both local and global anatomic structure influence the success of cardiological interventions, such as stent deployment or balloon angioplasty \cite{marlevi2021vascular,holm2023octbifstent,sawaya2016contemporarybif}. Local structure is characterized by the morphological attributes of each 3D component within the artery wall, such as size or shape \cite{otsuka2016pathologycoronary, mori2018coronaryprogressioncalcium}, while global structure is defined by the tree-like skeleton that results from the artery bifurcating into one or several branches.

Virtual intervention platforms aim to disentangle the complex relationship between coronary anatomy, biophysics, and pathophysiology \cite{stone2012predictionplaquemorph, doradla2020biomechanicalstressprofiling}, enabling physics-based simulations of medical interventions such as stent and balloon expansion \cite{poletti2022towardsadigitaltwin, karanasiou2020designstent,karanasiou2022silicotrial}. However, current simulation-based platforms face a trade-off between using anatomies that are either realistic or controllable. Using simplified parametric geometry \cite{dong2019impact} such as tubular cylinders to represent coronary arteries can isolate the effect of certain morphological features on device outcomes. However, this approach cannot capture the intricate 3D micro-morphology of coronary lesions \cite{virmani2006pathologycoronaryplaque,mori2018coronaryprogressioncalcium}, limiting the development of clinically relevant insights. On the other hand, reconstructing digital twins from medical images produces realistic anatomy but prevents control over anatomic attributes \cite{kadry2021platform,zhao2021patientbifurcationstent,
doradla2020biomechanicalstressprofiling}. As such, digital twins are unable to assess causal relationships between anatomic features and device effectiveness, as many of these features are correlated together. 

To address such limitations, we study how Latent Diffusion Models (LDMs) can be used as a controllable data source for comparative virtual intervention studies, producing customized anatomic counterfactuals that change the simulated outcome of device deployment. This would allow device designers to understand which subtle anatomic features influence interventional outcomes, guiding both the design process and clinical trial recruitment. We present three methodological adaptations that enable LDMs to synthesize and edit 3D multi-component coronary segmentation maps according to prespecified anatomic constraints. First, we study how to regularize LDM latent space to reduce the presence of topological defects (\cref{sec:topo_reg_loss}). Second, we enable conditional generation based on clinically interpretable and editable representations of morphology and skeletal structure (\cref{sec:morphoskeletal_cond}). The local morphological component is specified by one or several cross-sectional features along the length of the artery, while the global skeletal component specifies the presence, location, and shape of vessel bifurcations. Lastly, we extend current diffusion guidance strategies with morpho-skeletal regressor functions to enhance conditional fidelity (\cref{sec:morphoskeletal_guidance}). In addition, we introduce a novel guidance algorithm tailored to morphological conditioning which utilizes lightweight and non-differentiable morphological regressor functions to adaptively update the null condition during sampling (\cref{sec:morphoskeletal_guidance}). Our main contributions are thus as follows:
\begin{enumerate}
\item We introduce a method to improve the topological quality of 3D multi-component coronary anatomies produced by latent diffusion models through the use of a topological interaction loss during autoencoder training.
\item We develop a novel morpho-skeletal conditioning framework for latent diffusion models to enable the controlled synthesis of coronary arteries and extend loss-based guidance strategies to this task through the use of differentiable morpho-skeletal regressor functions.
\item We demonstrate the limitations of classifier-free guidance for morpho-skeletal generation and introduce an adaptive null guidance strategy. Our proposed method operates through the use of non-differentiable morphological regressors to determine the null guidance condition, enhancing conditional fidelity and computational efficiency.
\item We show that our framework can disentangle micro-morphology and arterial branch structure, condition on a variety of morphological features, and edit patient-specific coronary arteries for comparative simulation studies.
\end{enumerate}

\section{Related Work}
\label{sec:related}
\subsection{Generative Models of Virtual Anatomies}

Current generative models of coronary arteries are based on parametric approaches, utilizing geometric primitives such as cylinders to approximate coronary geometry \cite{dong2019impact,madani2019bridgingfemdeeplearningstress}. While such models enable geometric control, they do not capture the complex morphological or topological variation associated with coronary lesions. For simple topological organs such as the aorta, deep learning has been used to produce virtual anatomy by deforming a 3D template shape \cite{beetz2021generatingconditional,beetz2021generatingconditionalpointcloud,dou2022generativechimeras}. Template-based approaches have also been used for generative editing by applying localized elastic deformations to template shapes, inducing common vascular phenomena such as aneurysms and stenoses \cite{pham2023svmorph,pham2024virtualshape}. However, template-based approaches require anatomic correspondence and consistent topology between patients, which is not possible for multi-material coronary arteries. Alternatively, implicit representations \cite{qiao2022generativeageheart,qiao2023cheart,verhulsdonk2023shapeofmyheart,wiesner2023generativemodellivingcells}, enable flexible generation of topological structures but do not account for unphysiological topological defects. In contrast, we aim to generate implicit representations of coronary anatomy to enable flexible topological generation while selectively enforcing topological quality with topological losses.

\subsection{Diffusion Model Guidance}
Diffusion model guidance techniques primarily fall into two families. Loss-based methods include Classifier Guidance (CG) \cite{dhariwal2021diffusion}, which uses the gradient obtained from pre-trained classifiers to influence the sampling process. Loss-based techniques have also been extended to solve inverse problems, in which the correcting gradient is derived from a loss that measures the degree to which a constraint is satisfied \cite{chung2022diffusionposteriorsampling, he2023manifoldpreserving,song2023lossguided}. However, loss-based methods have not been studied in the context of anatomic constraints and require computationally expensive backpropagation. In contrast, null-conditioning based methods such as Classifier-Free Guidance (CFG) \cite{ho2022classifier} use a weighted combination of conditional and unconditional model outputs to guide the sampling process. Null conditions can also be defined in a negative manner, guiding the sampling process away from certain conditions \cite{gandikota2023erasingnegativeguide}. While effective for a variety of modalities \cite{nichol2021glide,ho2022classifier,wang2022semanticdiffusion}, the use of null-guidance for enforcing morphological or skeletal constraints has not been explored. In this study, we adapt diffusion guidance strategies to the task of enforcing morpho-skeletal constraints and develop novel guidance algorithms suited to conditioning with continuous morphological features.
\subsection{Topological regularization for Neural Networks}
\begin{wrapfigure}[17]{r}{0.5\textwidth}
    \vspace{-14mm}
  \begin{center}
    \includegraphics[width=0.5\textwidth]{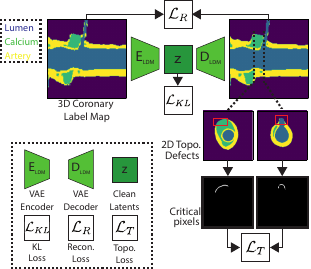}
  \end{center}
 \caption{\small We regularize the latent space $\mathbf{z}$ with a topological interaction loss applied to the calcium and lumen segmentations.}
  \label{fig:Method_VAE}
\end{wrapfigure}
Cardiovascular anatomy exhibits a wide variety of topological structures, manifesting as connected components, loops, and voids. It also features multi-tissue interactions such as containment and adjacency. Various techniques have been developed to constrain neural network outputs in accordance with prespecified topological priors. Persistent homology (PH) losses \cite{clough2020topologicalloss,byrne2022persistenthomologymulticlass} can regularize multiclass segmentation networks, where the individual classes and interactions between different classes should adhere to specified topological priors. However, such PH-based methods are prohibitively expensive to use with 3D data, especially as the number of topological interactions increase. Alternative techniques reduce topological interactions by penalizing critical voxels that violate topological priors \cite{gupta2022learningtopo}. We study the application of such strategies to diffusion models of anatomy, to which topological regularization has not been applied.

\section{Methodology}

\label{sec:methodology}

\subsection{Latent Diffusion Model}

We employ a latent diffusion model (LDM), which is trained in the latent space of a variational autoencoder (VAE). The training process for the VAE is shown in \cref{fig:Method_VAE} in which an encoder $E_{LDM}$ encodes the multi-tissue coronary segmentation maps $\mathbf{x} \in\mathbb{R}^{C\times H\times W\times D}$ into a latent representation $\mathbf{z} \in\mathbb{R}^{c\times h\times w\times d}$, where the latent dimensions are downsampled by a factor $f=H/h=W/w=D/d$. Here, $C$ is the number of tissue classes and $c$ is the number of latent channels. The latents are then passed to the decoder $D_{LDM}$ back into voxel space, where a Dice cross entropy reconstruction loss $\mathcal{L}_R$, KL divergence loss $\mathcal{L}_{KL}$, and a topological interaction loss $\mathcal{L}_T$ (\cref{sec:topo_reg_loss}) are applied. For our application, we aim to sample from the conditional probability distribution of encoded coronary segmentation maps $p_{data}(\mathbf{z}\vert \mathbf{y})$, where $\mathbf{y}$ represents mid-level anatomic constraints regarding morphology $\mathbf{y}_m$ and skeletal structure $\mathbf{y}_s$ (\cref{sec:morphoskeletal_cond}). To train a diffusion model, we consider the joint distribution $p(\mathbf{z}_\sigma\vert \mathbf{y};\sigma)$ obtained through a forward diffusion process, in which i.i.d Gaussian noise of standard deviation $\sigma$ is added to the data, where at $\sigma=\sigma_{max}$ the data is indistinguishable from Gaussian noise. Equivalently, the forward process can be described by the stochastic equation $d\mathbf{z}_\sigma =\sqrt{2\sigma}d\mathbf{w}$ where $\mathbf{w}$ is the standard Wiener process. The reverse diffusion process is defined as the solution to the following stochastic differential equation
\begin{equation}
d\mathbf{z}_\sigma = -  2\sigma \nabla_{\mathbf{z}_\sigma}\log p(\mathbf{z}_\sigma\vert \mathbf{y}; \sigma) \, dt+\sqrt{2\sigma}d\mathbf{w},
\label{eq:sde_back}
\end{equation}
Where $\nabla_{\mathbf{z}_\sigma}\log p(\mathbf{z_\sigma\vert \mathbf{y};\sigma})$ represents the  score function that is conditioned on $\mathbf{y}$. To train the network, we apply the forward diffusion process to produce intermediately noised latents $\mathbf{z}_\sigma=\mathbf{z}+\mathbf{n}$ where $\mathbf{n} \sim \mathcal{N}(\mathbf{0},\sigma^2\mathbf{I})$, parameterized by a noise level $\sigma$. The diffusion model is parameterized as a function $F_{\theta}$, encapsulated within a denoiser $D_{\theta}$, that takes the intermediately noised input $\mathbf{z}_\sigma$, the conditioning signal $\mathbf{y}$, and the noise level $\sigma$ to predict the clean data $\mathbf{z}$.
\begin{equation}
    D_{\theta}(\mathbf{z_\sigma};\sigma, \mathbf{y}) = 
    c_{\text{skip}}(\sigma)\,\mathbf{z_\sigma} + 
    c_{\text{out}}(\sigma)\,F_{\theta}(c_{\text{in}}(\sigma) \, \mathbf{z_\sigma}; c_{\text{noise}}(\sigma), \mathbf{y}) \,,
\end{equation}
where $c_{\text{skip}}$ allows $F_{\theta}$ to predict the noise $\mathbf{n}$ at low $\sigma$ and the training data $\mathbf{z}$ at high $\sigma$. The variables $c_{\text{out}}$ and $c_{\text{in}}$ scale the input and output magnitudes to be within unit variance, and the constant $c_{\text{noise}}$ maps the noise level $\sigma$ to a conditioning input for the network \cite{karras2022elucidating}. The denoiser output is related to the conditional score function through the relation $\nabla_{\mathbf{z}_\sigma} \log p(\mathbf{z}_\sigma\vert \mathbf{y}; \sigma) = \left( D_\theta(\mathbf{z_\sigma}; \sigma, \mathbf{y}) - \mathbf{z} \right) / \sigma^2$ and $F_{\theta}$ is chosen to be a 3D U-net (see \cref{appdx:Implementation}). The denoising loss $\mathcal{L}_D$ is then specified based on the agreement between the denoiser output and the original training data:
\begin{equation}
    \mathcal{L}_D=\mathbb E_{\sigma,\mathbf{z},\mathbf{n}}[\,\lambda(\sigma)\vert\vert D_\theta(\mathbf{z}_\sigma;\sigma, \mathbf{y})\,-\mathbf{z}\vert\vert^2_2] \,,
\end{equation}
such that the loss weighting $\lambda(\sigma)=1/c_{\text{out}}(\sigma)^2$ ensures an effective loss weight that is uniform across all noise levels, and $\sigma$ is sampled from a log-normal distribution.

\subsection{Topological Regularization}
To be compatible for numerical simulation, diseased tissue and lumen must be encompassed by the vessel wall within every 2D cross section of the artery. Non-compliance with these constraints can give rise to unrealistic biophysical phenomena. While post-processing can be applied to the segmentation map before meshing and simulation, such steps are time consuming and cumbersome, especially when done for each segmentation map \cite{straughan2023fully}. As such, we aim to enhance the topological viability of the coronary arteries sampled by the diffusion model. We study how to regularize the latent space of the autoencoder during training with a topological interaction loss, as described in Gupta et al. \cite{gupta2022learningtopo}. We adapt this loss to a 3D multi-slice context, using it to identify and penalize critical pixels in each 2D cross section with a cross-entropy loss function using vessel wall as a ground truth label (see \cref{appdx:Topo_loss}). The topological interaction loss  ($\mathcal{L}_{T}$) is then weighted and and added to the reconstruction ($\mathcal{L}_{R}$) and KL divergence ($\mathcal{L}_{KL}$) losses.

\label{sec:topo_reg_loss}

\begin{figure*}[h]
  \centering
  \includegraphics[width=1\textwidth]{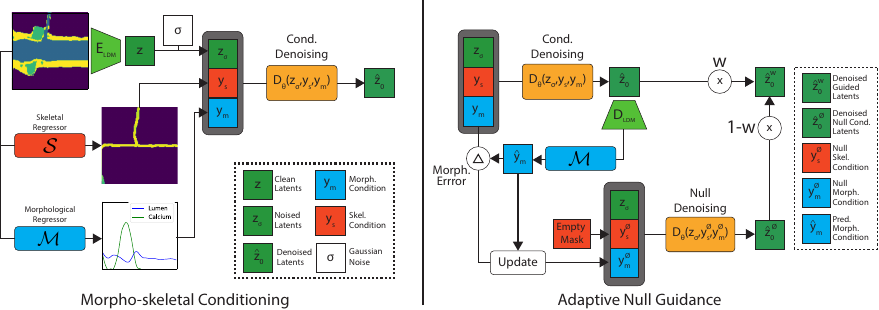}
  \caption{Left: We condition the training process of a latent diffusion model through channel-wise concatenation of the morphological ($\mathbf{y}_m$) and skeletal ($\mathbf{y}_s$) maps to the noised latent representation ($\mathbf{z}_\sigma$). Right: Our proposed guidance algorithm operates by specifying an null skeletal condition ($\mathbf{y}^\varnothing_s$)  and adaptively updating a null morphological condition ($\mathbf{y}^\varnothing_m$) based on the error between the target condition ($\mathbf{y}_m$) and the morphology exhibited by the current denoiser output ($\mathbf{\hat{y}}_m$).}
  \label{fig:Method_DDM}
\end{figure*}

\subsection{Morpho-skeletal Conditioning}
\label{sec:morphoskeletal_cond}
\subsubsection{Morphological Conditioning}
The morphological conditioning process is shown in \cref{fig:Method_DDM} (left). We use a morphological regressor $\mathcal{M}$ to compute a morphological conditioning map $\mathbf{y}_m =\mathcal{M}(\mathbf{x}) \in \mathbb{R}^{m \times h \times w \times d}$ that is concatenated to the latent representation $\mathbf{z}$ to condition the denoising process. To do this, our morphological regressor $\mathcal{M}$ spatially encodes local morphological information through cross sectional features along the depth dimension of the straightened coronary segmentation map. This process results in a set of morphological feature vectors $\mathbf{M} \in \mathbb{R}^{m \times D} $, where $m$ represents the number of 2D morphological attributes and $D$  signifies the number of frames along the artery centerline. The features are calculated, normalized by the 2nd and 98th percentile attribute values within the training set, and appropriately downsampled for concatenation (see \cref{appdx:Morph_Features}).

\subsubsection{Skeletal Conditioning}
The skeletal conditioning process is shown in \cref{fig:Method_DDM} (left). We use a skeletal regressor  $\mathcal{S}$ to compute a skeletal conditioning map $\mathbf{y}_s =\mathcal{S}(\mathbf{x}) \in \mathbb{R}^{1 \times h \times w \times d}$ that is concatenated to the latent representation $\mathbf{z}$ to condition the denoising process. To do this, our skeletal regressor $\mathcal{S}$ encodes global bifurcational information by finding the luminal skeleton $ \mathbf{S} \in \mathbb{R}^{H \times W \times D}$, which describes the branching structure of the coronary arteries. We then compute, process, and resize the skeleton map to produce the skeletal conditioning signal (see \cref{appdx:Skel_Features}). We zero out the skeletal condition with a probability of 0.2 to jointly train an unconditional model with respect to the lumen skeleton.

\subsection{Morpho-skeletal Guidance}
\label{sec:morphoskeletal_guidance}
\subsubsection{Loss Guidance}
We formulate loss-based guidance in terms of the gradient derived from a morphological loss $\mathcal{L}_m$ and skeletal loss $\mathcal{L}_s$, as can be seen in \cref{alg:DPS}. Each loss term is defined as the L2 norm between the target morpho-skeleton and the morpho-skeleton measured from synthetic segmentation maps by the differentiable regressor functions $\bar{\mathcal{M}}$ and  $\bar{\mathcal{S}}$ (see \cref{appdx:Regressors}). 
\begin{equation}
    \mathcal{L}_m=\lVert \mathbf{y}_m-\bar{\mathcal{M}}(\mathbf{x})\rVert_2^2  \quad\mathrm{and}\quad 
        \mathcal{L}_s=\lVert \mathbf{y}_s-\bar{\mathcal{S}}(\mathbf{x})\rVert_2^2,
\end{equation}
where the default choice of segmentation maps $\mathbf{x}$ is obtained by decoding the denoised latent from the diffusion model $\mathbf{\hat{z}}_0$ such as that in Chung et al. \cite{chung2022diffusionposteriorsampling}. Classifier guidance \cite{ho2022classifier} can be implemented by decoding the intermediately noised latents $\mathbf{z}_\sigma$ to obtain $\mathbf{x}$. Finally, we guide the conditional denoising process by using the gradient of the loss with respect to $\mathbf{z_\sigma}$ and a guidance term $(w-1):$
\begin{equation}
\scalebox{0.9}{ 
$\underbrace{D_{\theta}^w (\mathbf{z_\sigma};\sigma, \mathbf{y}_m,\mathbf{y}_s)}_{\text{Guided Denoising}} = \underbrace{D_{\theta}(\mathbf{z_\sigma};\sigma,\mathbf{y}_m,\mathbf{y}_s)}_{\text{Cond. Denoising}}+\underbrace{(w-1) \nabla_{\mathbf{z_\sigma}}  (\mathcal{L}_m+\mathcal{L}_s)}_{\text{Gradient Guidance}},$
}
\end{equation}
\subsubsection{Adaptive Null guidance}
To circumvent the computational cost of backpropagating through the diffusion model, VAE decoder, and regressors, we formulate a null-conditional guidance method based on simple and non differentiable morphological measurement functions, as can be seen in \cref{fig:Method_DDM} (right) and \cref{alg:ANG}. We take inspiration from classifier-free guidance (CFG), which guides the denoising process by a weighted combination of the denoiser output conditioned by the target conditioning signals $(\mathbf{y}_m,\mathbf{y}_s)$ and the null conditioning signals
$(\mathbf{y}^\varnothing_m, \mathbf{y}^\varnothing_s)$:

\begin{equation}
\scalebox{0.9}{ 
$ \underbrace{D_{\theta}^w (\mathbf{z_\sigma};\sigma, \mathbf{y}_m,\mathbf{y}_s)}_{\text{Guided Denoising}} =\underbrace{w D_{\theta}(\mathbf{z_\sigma};\sigma, \mathbf{y}_m,\mathbf{y}_s)}_{\text{Cond. Denoising}}+ \underbrace{(1-w) D_{\theta}(\mathbf{z_\sigma};\sigma, \mathbf{y}^\varnothing_m, \mathbf{y}^\varnothing_s)}_{\text{Null Denoising}}, $
}
\end{equation}

where $w$ is a guidance weight typically set to be larger than 1. To enable conventional classifier-free guidance, we train an unconditional diffusion model and use it to guide sampling. We implement our proposed algorithm, adaptive null guidance (ANG), by adaptively updating the null signal for the morphological condition $\mathbf{y}^\varnothing_{m}$ based on the difference between the target morphology $\mathbf{y}_{m}$ and the morphology  derived from the current denoised output $\mathbf{\hat{y}}_{m}=\mathcal{M}(\mathbf{\hat{x}}_0)$.

\begin{equation}
\mathbf{y}^\varnothing_{\text{m}}=\mathcal{M}(\mathbf{\hat{x}}_0)+\underbrace{(\mathcal{M}(\mathbf{\hat{x}}_0)-\mathbf{y}_m)}_{\text{Morph. Error}},
\end{equation}
where $\mathcal{M}$ is a morphological regressor function that is not required to be differentiable, and the segmentation map $\mathbf{\hat{x}}_0$ is decoded from the denoiser output $\mathbf{\hat{z}}_0=D_\theta(\mathbf{z_\sigma},\mathbf{y}_m,\mathbf{y}_s,\sigma)$. Skeletal guidance is implemented similar to conventional CFG, where the skeletal condition $\mathbf{y}^\varnothing_{s}$ is set as an empty mask.

\subsubsection{Morpho-skeletal Regressors for Guidance}
Our proposed guidance methods require morphological or skeletal regressors to guide the reverse diffusion process. For null guidance, we leverage nondifferentiable morphological regressor functions $\mathcal{M}$, as well as a hard skeletonization method $\mathcal{S}$, developed by Sato et al. \cite{sato2000teasar}, which produces high fidelity estimations of the coronary morphoskeleton. Loss guidance, in contrast, requires that the regressors be differentiable to calculate gradients. A soft morphological regressor $\mathcal{\bar{M}}$, consisting of differentiable image processing operations, is used when possible for morphological loss guidance. We further implement a neural morphological regressor  $\bar{\mathcal{M}}_\phi$, to extend loss guidance to morphological features that cannot be easily derived in a differentiable fashion. The neural morphological regressor is trained on high fidelity morphological features from the ground truth morphological regressor $\mathcal{M}$. For skeletal loss guidance, our skeletal regressor $\bar{\mathcal{S}}$ leverages an iterative soft skeletonization algorithm \cite{shit2021cldice} consisting of erosions and dilations. However, since soft skeletonization methods can exhibit limited performance compared to non-differentiable methods, we also implement a neural skeletal regressor  $\bar{\mathcal{S}}_\psi$ to regress high fidelity skeletons (see Appendix \ref{appdx:Regressors}).

\begin{minipage}[t]{0.48\textwidth}
\begin{algorithm}[H]
\caption{Loss Guidance}
\begin{algorithmic}[1]
\Require $D_\theta, \mathbf{y}_m,\mathbf{y}_s,\sigma, \mathbf{z}_\sigma, w,\bar{\mathcal{M}}, \bar{\mathcal{S}}$ 
\State $\hat{\mathbf{z}}_0 \gets D_\theta(\mathbf{z}_\sigma; \sigma,\mathbf{y}_m,\mathbf{y}_s)$

\State $\hat{\mathbf{x}}_0 \gets D_{LDM}(\hat{\mathbf{z}}_0 )$

\State $\mathcal{L}_m \gets \lVert \mathbf{y}_m-\bar{\mathcal{M}}(\hat{\mathbf{x}}_0)\rVert_2^2$

\State $\mathcal{L}_s \gets \lVert\mathbf{y}_s-\bar{\mathcal{S}}(\hat{\mathbf{x}}_0)\rVert_2^2$

\State $\hat{D}_\theta\gets \hat{\mathbf{z}}_0+(w-1)\nabla_{\mathbf{z}_\sigma}(\mathcal{L}_m+\mathcal{L}_s)$

\State \textbf{return} $\hat{D}_\theta$

\end{algorithmic}
\label{alg:DPS}
\end{algorithm}
\end{minipage}
\hfill
\begin{minipage}[t]{0.48\textwidth}
\begin{algorithm}[H]
\caption{Adaptive Null Guidance}
\begin{algorithmic}[1]
\Require $D_\theta, \mathbf{y}_m,\mathbf{y}_s,\sigma, \mathbf{z}_\sigma, w, \mathcal{M}$ 
\State $\hat{\mathbf{z}}_0 \gets D_\theta(\mathbf{z}_\sigma; \sigma,\mathbf{y}_m,\mathbf{y}_s)$

\State $\hat{\mathbf{x}}_0 \gets D_{LDM}(\hat{\mathbf{z}}_0 )$

\State $\mathbf{y}_m^\varnothing \gets \mathcal{M}(\hat{\mathbf{x}}_0)+(\mathcal{M}(\hat{\mathbf{x}}_0)-\mathbf{y}_m)$

\State $\mathbf{y}_s^\varnothing \gets \varnothing$

\State $\hat{D}_\theta\gets w\hat{\mathbf{z}}_0+(1-w)D_\theta(\mathbf{z}_\sigma; \sigma,\mathbf{y}_m^\varnothing,\mathbf{y}_s^\varnothing)$

\State \textbf{return} $\hat{D}_\theta$
\end{algorithmic}
\label{alg:ANG}
\end{algorithm}
\end{minipage}

\section{Experiments}

\subsection{Dataset}
We employ a dataset comprising 62 Coronary Computed Tomography (CCTA) images of patients with coronary artery disease \cite{sonck2022clinical}, from which 222 unique 3D segmentation map patches were extracted (see \cref{appdx:Dataset}). Each segmentation map contains lumen, vessel wall, and diseased calcified tissue within the straightened coronary artery. Consequently, each 3D segmentation map has dimensions of $4\times 128 \times 128 \times 128$. As the lateral resolution of the segmentation maps is larger than the in-plane resolution, the arterial segmentation maps in the results section are visualized with isotropic resolutions by interpolation along the depth dimension.

\subsection{Evaluation Metrics}
We calculate conditional fidelity, morphological quality, and topological quality for evaluation. Conditional fidelity is measured for both morphological and skeletal conditioning. Morphological conditioning fidelity is defined as the mean absolute error between the target morphological features and the features derived from the synthetic coronary segmentation map. Skeletal conditioning fidelity is defined as the mean absolute error between the number of branches derived from the target skeleton and the skeleton derived from the synthetic coronary segmentation map. To quantify morphological quality, we record 12 morphological features of each 3D artery in addition to 8 features for all 2D cross sections (see \cref{appdx:Morph_Features}). We calculate the Frechet Distance (FD), Precision, and Recall between generated and training segmentation maps for both 3D and 2D features respectively, similar to previous approaches \cite{kadry2023probing}. For topological quality, we define the \% of topological violations as the proportion of 2D cross sections within each coronary segmentation map that contains a critical pixel, defined by adjacency relations between the calcium, lumen, or background labels.

\subsection{Enforcing Topological Constraints}
\begin{figure}[h]
  \centering
   \includegraphics[width=0.8\linewidth]{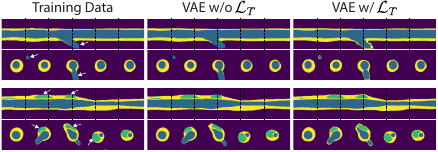}
   \caption{Longitudinal and cross sectional slices of coronary segmentation maps. First column shows that patient-specific segmentation maps from the training set exhibit several topological defects such as lumen and calcium tissues not being fully contained within vessel wall tissue (white arrows). Second and third columns show the effect of autoencoding coronary segmentation maps with (right) or without (center) using a topological loss $\mathcal{L}_T$ during training.}
   \label{fig:A_topo_violations}
\end{figure}

We first compare the effect of applying our topological loss to the VAE latent space. We train two VAE models with and without a topological loss and use them to reconstruct 100 coronary segmentation maps in our training set. We specifically enforce that the calcium and lumen tissues be encompassed by the vessel wall. \cref{VAE_Topo} shows that VAEs reconstruct arteries with similar levels of topological errors to the train set. When using a topological loss during training, the rate of topological errors is reduced by an order of magnitude.  This can be seen in \cref{fig:A_topo_violations} where topological regularization reduces the presence of topological defects. We then compare the effect of topologically regularized VAEs on LDM training. We train two unconditional LDM's and sample 200 coronary artery segmentation maps from each model, using 100 sampling steps each. \cref{VAE_Topo} shows topological regularization of the VAE latent space reduces topological violations in synthetic segmentation maps.

\begin{table}[t]
\centering
\caption{Left: Topological interaction violations exhibited by segmentation maps that were autoencoded by various VAE models. Training a VAE with topological interaction losses improves topological quality. Right: Topological interaction violations by segmentation maps sampled from LDMs. Topological regularization of the LDM latent space improves the topological quality of generated arteries.}

\begin{minipage}[t]{.4\linewidth}
\centering
\scalebox{1}{%
\setlength{\tabcolsep}{2pt} 
\begin{tabular}{c|cc}
\hline
\multicolumn{1}{c|}{VAE} & \multicolumn{2}{c}{Topo. Viol. (\%)} \\ \cline{2-3}
Training & Lumen & Calcium \\ \hline
Train Set & 18 & 16\\
$\mathcal{L}_R+\mathcal{L}_{KL}$ & 17 & 15 \\ 
$\mathcal{L}_R+\mathcal{L}_{KL}+ \mathcal{L}_T$& \textbf{0.5} & \textbf{1.1} \\ 
\hline
\end{tabular}
}
\label{VAE_Topo}
\end{minipage}%
\hfill
\begin{minipage}[t]{.5\linewidth}
\centering
\scalebox{1}{%
\setlength{\tabcolsep}{4pt} 
\begin{tabular}{c|cc}
\hline
\multicolumn{1}{c|}{LDM} & \multicolumn{2}{c}{Topo. Viol. (\%)} \\ \cline{2-3}
Training & Lumen & Calcium \\ \hline
VAE w/o $\mathcal{L}_T$ & 1.44 & 9.21 \\ 
VAE w $\mathcal{L}_T$ & \textbf{0.10} & \textbf{2.61} \\ 
\hline
\end{tabular}
}
\label{LDM_Topo}
\end{minipage}
\label{table:combined}
\end{table}

\subsection{Enforcing Morpho-skeletal Constraints}

\begin{figure}[t]
  \centering
  \includegraphics[width=0.9\linewidth]{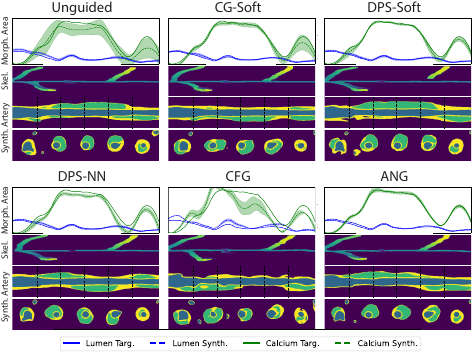}
  \caption{Example morphological features, skeleton depth maps and synthetic segmentation map cross sections for various guidance methods. Filled in regions within morphological plots indicate standard deviation over 10 generated segmentation maps. A guidance weight of 5 is used when applicable. Our proposed guidance method (ANG) improves conditional fidelity while maintaining good visual quality.}
  \label{fig:Guidance_Labelmap}
\end{figure}

\begin{figure}[h]
  \centering
  \includegraphics[width=\linewidth]{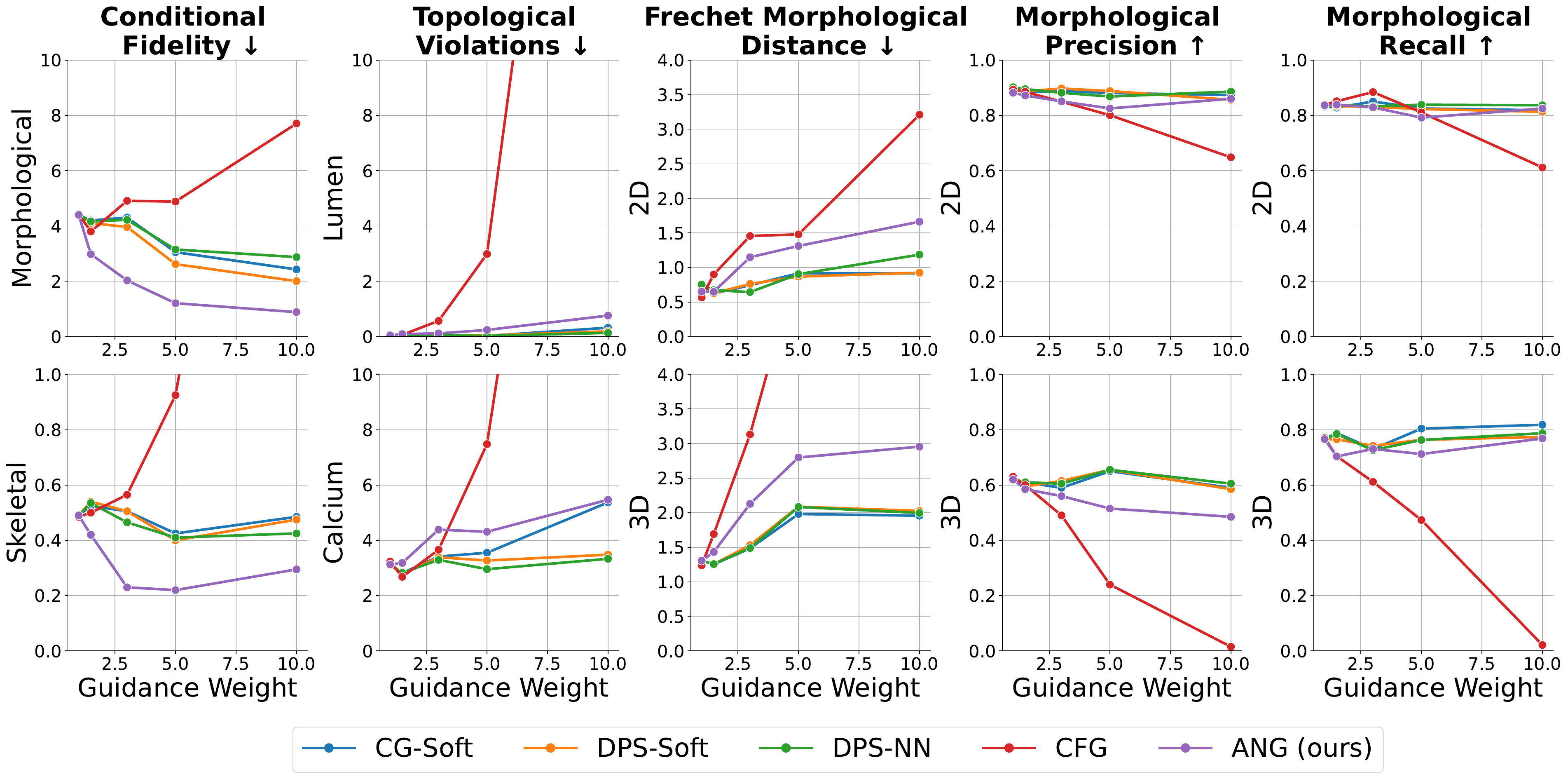}
  \caption{Conditional fidelity and quality metrics for various guidance methods over several guidance weights. Our guidance method (ANG) enables enhanced conditioning fidelity at the cost of slightly decreased morphological and topological quality.}
  \label{fig:Guidance}
\end{figure}

\subsubsection{Conditioning Signal Ablation Study}
\label{Cond_signal_ablation_sec}
We examine the impact of morphological and skeletal conditioning on conditional fidelity and morphological quality in \cref{table:gpu_comparison}. To condition only on morphology or skeletal structure, we train the original LDM and drop the corresponding conditioning signal. For consistency, our unconditional model is trained by dropping out both conditions. Our default morphological condition consists of the cross-sectional area curves for the lumen and calcium tissues. The morphological and skeletal features were derived from the validation set and used to sample 200 segmentation maps with 100 sampling steps. Joint conditioning of both skeleton and morphology shows similar conditioning fidelity to conditioning on each constraint alone. Morphological conditioning alone improves skeletal fidelity as compared to unconditional sampling, likely due to the lumen area curve weakly indicating the presence of bifurcations.

\subsubsection{Evaluating Morpho-skeletal Guidance Strategies} We study the impact of varying the morpho-skeletal guidance weight on a morpho-skeletally conditioned LDM for various guidance methods. We evaluate three strategies for loss guidance and two strategies for null guidance.  For loss guidance, we implement i) classifier-guidance (CG-Soft) and ii) diffusion posterior sampling with differentiable morpho-skeletal regressors $\bar{\mathcal{M}}$ and $\bar{\mathcal{S}}$ (DPS-Soft)  as well as iii) a variant of DPS with neural network regressors $\bar{\mathcal{M}}_\phi$ and $\bar{\mathcal{S}}_\psi$ (DPS-NN). For null guidance, we implement i) classifier-free guidance (CFG) using a separately trained unconditional model and ii) Adaptive Null-Guidance (ANG) with a non-differentiable morphological regressor $\mathcal{M}$. We evaluated every method by sampling 200 arteries with 25 sampling steps for each combination of guidance weight and method. The input morpho-skeletal conditions were sampled from the validation set. Qualitative results can be seen in \cref{fig:Guidance_Labelmap}, where an unguided diffusion model can capture the variation in the lumen area but is unable to achieve high fidelity with respect to the calcium area. We see that morpho-skeletal guidance enhances conditional fidelity for all methods but CFG, which produces low quality segmentation maps with increasing guidance. ANG exhibits the best mix of conditioning fidelity and visual quality as compared to a method such as DPS-Soft, which has good conditional fidelity but produces un-physiological artifacts. These observations are reinforced in \cref{fig:Guidance}, which shows that ANG outperforms all methods in terms of both morphological and skeletal conditioning fidelity, while maintaining similar or slightly decreased recall values. Furthermore, ANG exhibits the lowest sampling time and memory usage for regressor based guidance strategies (\cref{table:gpu_comparison}) as it does not require the use of backpropagation. However, it can be seen that increasing the guidance weight for ANG increases the number of topological violations and causes a decrease in 3D morphological precision. 

\begin{figure}[h]
  \centering
  \includegraphics[width=\linewidth]{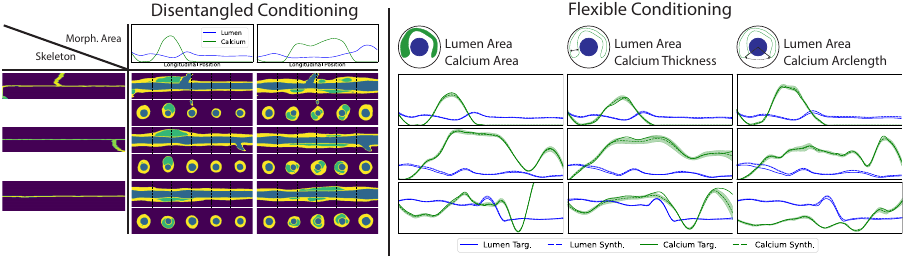}
  \caption{Left: Synthetic coronary segmentation maps conditioned on different morphological features (columns) and lumen skeletons (rows) obtained from patient specific coronary segmentation maps. Right: Morphological features produced by conditioning on different combinations of lumen and calcium features (columns). We used adaptive null guidance with a weight of 5 to sample all coronary segmentation maps.}
  \label{fig:B_Disentangled}
\end{figure}

\begin{table}[h]
\centering
\caption{Left: Ablation study for conditioning mechanisms. Joint conditioning improves conditional fidelity and morphological quality as compared to morphological conditioning alone. Right: Sampling speed and GPU memory usage comparison for a single sample over different guidance methods with 100 sampling steps. ANG is the most computationally efficient regressor-based guidance method.}
\begin{minipage}{.55\textwidth}
\centering

\begin{tabular}{lcc|cc|cc}
\hline
& \multicolumn{2}{c}{\textbf{Conditional}} & \multicolumn{2}{c}{\textbf{Morph.}} & \multicolumn{2}{c}{\textbf{Morph.}} \\
\textbf{Method} & \multicolumn{2}{c}{\textbf{Fidelity}} & \multicolumn{2}{c}{\textbf{Prec.}} & \multicolumn{2}{c}{\textbf{Rec.}} \\
\textbf{} & Morph. & Skel. & 3D & 2D & 3D & 2D\\ \hline
Uncond. & 117 & 3.2 & \textbf{0.68} & 0.89 & \textbf{0.85} & 0.72 \\
Morph. & \textbf{4.0} & 1.2 & 0.65 & 0.88 & \textbf{0.93} & 0.82 \\
Skel. & 117 & \textbf{0.47} & 0.62 & 0.88 & 0.79 & 0.73 \\
Joint. & 4.5 & 0.48 & 0.64 & \textbf{0.89} & 0.79 & \textbf{0.86} \\
\hline
\end{tabular}
\label{Tab:Ablation_Cond}

\end{minipage}%
\begin{minipage}{.45\textwidth}
\centering
\begin{tabular}{c c c}
\hline
\multicolumn{1}{c}{Guidance} & \multicolumn{1}{c}{Sample} & \multicolumn{1}{c}{Memory} \\
\multicolumn{1}{c}{Method} & \multicolumn{1}{c}{Time (s)} & \multicolumn{1}{c}{(GB)} \\
\hline
\multicolumn{1}{c}{\textbf{No Regressor}} \\
\hline
Unguided & \textbf{3} & \textbf{2.35} \\
CFG & 9 & 3.07 \\
\hline
\multicolumn{1}{c}{\textbf{Regressor}} \\
\hline
CG-Soft & 66 & 12.5\\
DPS-Soft & 89 & 15.1 \\
DPS-NN & 80 & 8.6 \\
ANG (ours) & \textbf{46} & \textbf{6.4} \\
\hline
\end{tabular}
\end{minipage}

\label{table:gpu_comparison}
\end{table}

\subsection{Applications}
\subsubsection{Disentangled Morpho-skeletal Conditioning}
To demonstrate disentangled control over both coronary morphology and luminal skeleton, we use two sets of morphological feature vectors $\mathbf{M}$ and three skeleton heatmaps $\mathbf{S}$ derived from patient-specific arteries. We then sample 6 synthetic segmentation maps that correspond to all conditioning combinations. \cref{fig:B_Disentangled} shows that the same morphological conditioning signal can manifest differently based on the skeletal conditioning. For example, the combination of a local rise in lumen area in tandem with a skeletal branch induces a bifurcation in the generated model, but given a branch-less skeleton, the local rise in area manifests as an increase in lumen area.

\subsubsection{Flexible Morpho-skeletal Conditioning}
Our framework enables morphological conditioning using a variety of features, we train our LDM on different combinations of 1) lumen area and 2) a morphological feature belonging to calcium, sampling 10 segmentation maps each. We specifically condition on calcium area, maximum calcium thickness, and maximum calcium arclength, using non differentiable morphological regressor functions. \cref{fig:B_Disentangled} demonstrates our guidance mechanism can enforce a variety of morphological constraints with high fidelity, without the need for training neural morphological regressors.

\subsubsection{Editing with Morpho-skeletal Conditioning}
\begin{figure*}[h]
  \centering
  \includegraphics[width=0.9\textwidth]{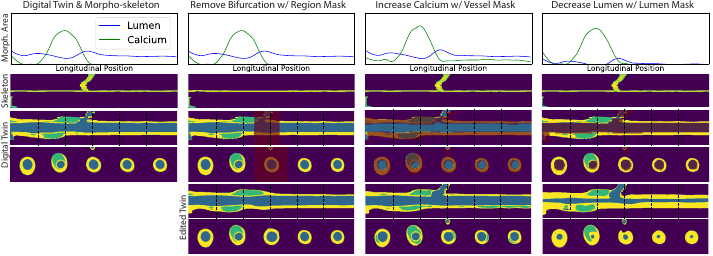}
  \caption{Examples of morpho-skeleton based editing procedures applied to a coronary digital twin (left). The patient-specific morpho-skeleton is edited and combined with a spatial mask (red) to inpaint clinically relevant features into the artery.}
  \label{fig:editing}
\end{figure*}

To virtually edit a coronary artery segmentation, we specify three components: morphological feature vectors, a skeleton heatmap, and a corresponding mask. We either specify region based masks, which mask out entire sections along the length of the artery, or tissue based masks, which mask out all voxels of pre-specified labels. We edit a coronary digital twin to create three different edited variants (\cref{fig:editing}). The first edit eliminates the bifurcation within the artery, achieved through regional editing with a branch-less skeleton heatmap. The second edit increases the amount of calcium by modifying the morphological feature vectors and applying a tissue mask to maintain the boundaries of the vessel wall. The third edit results in luminal narrowing by reducing the lumen area through the morphological feature vectors and employing a dilated tissue mask over the lumen.
\label{sec:editing}
\section{Limitations}
\label{sec:impact}

Our method has several limitations. First, our morphological conditioning mechanism can only capture morphological features along a pre-defined direction, limiting its applicability. Second, the use of 3D voxels confines our method to low dimensional segmentation maps. Third, our guidance strategies requires the use of the decoder for every sampling step, increasing computational costs. Lastly, our topological regularization method reduces but does not eliminate topological errors, meaning that arteries sampled from our model must still undergo some amount topological post-processing.

\section{Conclusions}
\label{sec:conclusions}
We adapt diffusion models to conditionally generate or edit coronary atherosclerotic anatomy that can be used for comparative numerical simulation studies of coronary interventions. We introduce several methodological changes to the training process of latent diffusion models to enable precise control over topological quality, local micro-morphology, and global skeletal structure. We further compare and contrast different guidance strategies to ensure adherence to morpho-skeletal constraints, proposing a novel and lightweight guidance method that can utilize non-differentiable regressor functions to guide sampling. Our anatomic generation framework offers the flexibility and control of simplified parametric geometries while maintaining the realism and 3-dimensional complexity of patient-specific geometries derived from medical images.

\section*{Acknowledgments}
We thank Vivek Gopalakrishnan, Neel Dey, Neerav Karani, Anurag Vaidya, and Ajay Manicka for discussion and comments. Funding for this project was provided by the National Institutes of Health (GM 49039) and Shockwave Medical.
\bibliographystyle{splncs04}
\bibliography{egbib}

\begin{thebibliography}{10}
\providecommand{\url}[1]{\texttt{#1}}
\providecommand{\urlprefix}{URL }
\providecommand{\doi}[1]{https://doi.org/#1}

\bibitem{beetz2021generatingconditional}
Beetz, M., Banerjee, A., Grau, V.: Generating subpopulation-specific biventricular anatomy models using conditional point cloud variational autoencoders. In: International Workshop on Statistical Atlases and Computational Models of the Heart. pp. 75--83. Springer (2021)

\bibitem{beetz2021generatingconditionalpointcloud}
Beetz, M., Banerjee, A., Grau, V.: Generating subpopulation-specific biventricular anatomy models using conditional point cloud variational autoencoders. In: International Workshop on Statistical Atlases and Computational Models of the Heart. pp. 75--83. Springer (2021)

\bibitem{byrne2022persistenthomologymulticlass}
Byrne, N., Clough, J.R., Valverde, I., Montana, G., King, A.P.: A persistent homology-based topological loss for cnn-based multiclass segmentation of cmr. IEEE transactions on medical imaging  \textbf{42}(1),  3--14 (2022)

\bibitem{chung2022diffusionposteriorsampling}
Chung, H., Kim, J., Mccann, M.T., Klasky, M.L., Ye, J.C.: Diffusion posterior sampling for general noisy inverse problems. arXiv preprint arXiv:2209.14687  (2022)

\bibitem{clough2020topologicalloss}
Clough, J.R., Byrne, N., Oksuz, I., Zimmer, V.A., Schnabel, J.A., King, A.P.: A topological loss function for deep-learning based image segmentation using persistent homology. IEEE transactions on pattern analysis and machine intelligence  \textbf{44}(12),  8766--8778 (2020)

\bibitem{abaqus2023}
{Dassault Systèmes}: Abaqus {F}inite {E}lement {A}nalysis {S}oftware. Vélizy-Villacoublay, France, version 2023 edn. (2023), available from Dassault Systèmes, https://www.3ds.com

\bibitem{dhariwal2021diffusion}
Dhariwal, P., Nichol, A.: Diffusion models beat gans on image synthesis. Advances in neural information processing systems  \textbf{34},  8780--8794 (2021)

\bibitem{dong2019impact}
Dong, P., Bezerra, H.G., Wilson, D.L., Gu, L.: Impact of calcium quantifications on stent expansions. Journal of biomechanical engineering  \textbf{141}(2),  021010 (2019)

\bibitem{doradla2020biomechanicalstressprofiling}
Doradla, P., Otsuka, K., Nadkarni, A., Villiger, M., Karanasos, A., Zandvoort, L.J.v., Dijkstra, J., Zijlstra, F., Soest, G.v., Daemen, J., et~al.: Biomechanical stress profiling of coronary atherosclerosis: identifying a multifactorial metric to evaluate plaque rupture risk. Cardiovascular imaging  \textbf{13}(3),  804--816 (2020)

\bibitem{dou2022generativechimeras}
Dou, H., Virtanen, S., Ravikumar, N., Frangi, A.F.: A generative shape compositional framework: Towards representative populations of virtual heart chimaeras. arXiv preprint arXiv:2210.01607  (2022)

\bibitem{gandikota2023erasingnegativeguide}
Gandikota, R., Materzynska, J., Fiotto-Kaufman, J., Bau, D.: Erasing concepts from diffusion models. In: Proceedings of the IEEE/CVF International Conference on Computer Vision. pp. 2426--2436 (2023)

\bibitem{gupta2022learningtopo}
Gupta, S., Hu, X., Kaan, J., Jin, M., Mpoy, M., Chung, K., Singh, G., Saltz, M., Kurc, T., Saltz, J., et~al.: Learning topological interactions for multi-class medical image segmentation. In: European Conference on Computer Vision. pp. 701--718. Springer (2022)

\bibitem{he2023manifoldpreserving}
He, Y., Murata, N., Lai, C.H., Takida, Y., Uesaka, T., Kim, D., Liao, W.H., Mitsufuji, Y., Kolter, J.Z., Salakhutdinov, R., et~al.: Manifold preserving guided diffusion. arXiv preprint arXiv:2311.16424  (2023)

\bibitem{ho2022classifier}
Ho, J., Salimans, T.: Classifier-free diffusion guidance. arXiv preprint arXiv:2207.12598  (2022)

\bibitem{holm2023octbifstent}
Holm, N.R., Andreasen, L.N., Neghabat, O., Laanmets, P., Kumsars, I., Bennett, J., Olsen, N.T., Odenstedt, J., Hoffmann, P., Dens, J., et~al.: Oct or angiography guidance for pci in complex bifurcation lesions. New England Journal of Medicine  \textbf{389}(16),  1477--1487 (2023)

\bibitem{kadry2023probing}
Kadry, K., Gupta, S., Nezami, F.R., Edelman, E.R.: Probing the limits and capabilities of diffusion models for the anatomic editing of digital twins. arXiv preprint arXiv:2401.00247  (2023)

\bibitem{kadry2021platform}
Kadry, K., Olender, M.L., Marlevi, D., Edelman, E.R., Nezami, F.R.: A platform for high-fidelity patient-specific structural modelling of atherosclerotic arteries: from intravascular imaging to three-dimensional stress distributions. Journal of the Royal Society Interface  \textbf{18}(182),  20210436 (2021)

\bibitem{karanasiou2020designstent}
Karanasiou, G.S., Tsobou, P.I., Tachos, N.S., Antonini, L., Petrini, L., Pennati, G., Gijsen, F., Nezami, F.R., Tzafiri, R., Vaughan, T., et~al.: Design and implementation of in silico clinical trial for bioresorbable vascular scaffolds. In: 2020 42nd Annual International Conference of the IEEE Engineering in Medicine \& Biology Society (EMBC). pp. 2675--2678. IEEE (2020)

\bibitem{karanasiou2022silicotrial}
Karanasiou, G.S., Tsompou, P.I., Tachos, N., Karanasiou, G.E., Sakellarios, A., Kyriakidis, S., Antonini, L., Pennati, G., Petrini, L., Gijsen, F., et~al.: An in silico trials platform for the evaluation of stent design effect in post-implantation outcomes. In: 2022 44th Annual International Conference of the IEEE Engineering in Medicine \& Biology Society (EMBC). pp. 4970--4973. IEEE (2022)

\bibitem{karras2022elucidating}
Karras, T., Aittala, M., Aila, T., Laine, S.: Elucidating the design space of diffusion-based generative models. arXiv preprint arXiv:2206.00364  (2022)

\bibitem{madani2019bridgingfemdeeplearningstress}
Madani, A., Bakhaty, A., Kim, J., Mubarak, Y., Mofrad, M.R.: Bridging finite element and machine learning modeling: stress prediction of arterial walls in atherosclerosis. Journal of biomechanical engineering  \textbf{141}(8),  084502 (2019)

\bibitem{marlevi2021vascular}
Marlevi, D., Edelman, E.R.: Vascular lesion--specific drug delivery systems: Jacc state-of-the-art review. Journal of the American College of Cardiology  \textbf{77}(19),  2413--2431 (2021)

\bibitem{mori2018coronaryprogressioncalcium}
Mori, H., Torii, S., Kutyna, M., Sakamoto, A., Finn, A.V., Virmani, R.: Coronary artery calcification and its progression: what does it really mean? JACC: Cardiovascular Imaging  \textbf{11}(1),  127--142 (2018)

\bibitem{myronenko20193dsegresnet}
Myronenko, A.: 3d mri brain tumor segmentation using autoencoder regularization. In: Brainlesion: Glioma, Multiple Sclerosis, Stroke and Traumatic Brain Injuries: 4th International Workshop, BrainLes 2018, Held in Conjunction with MICCAI 2018, Granada, Spain, September 16, 2018, Revised Selected Papers, Part II 4. pp. 311--320. Springer (2019)

\bibitem{nichol2021glide}
Nichol, A., Dhariwal, P., Ramesh, A., Shyam, P., Mishkin, P., McGrew, B., Sutskever, I., Chen, M.: Glide: Towards photorealistic image generation and editing with text-guided diffusion models. arXiv preprint arXiv:2112.10741  (2021)

\bibitem{otsuka2016pathologycoronary}
Otsuka, F., Yasuda, S., Noguchi, T., Ishibashi-Ueda, H.: Pathology of coronary atherosclerosis and thrombosis. Cardiovascular diagnosis and therapy  \textbf{6}(4), ~396 (2016)

\bibitem{pham2024virtualshape}
Pham, J., Kong, F., James, D.L., Marsden, A.L.: Virtual shape-editing of patient-specific vascular models using regularized kelvinlets. IEEE Transactions on Biomedical Engineering  (2024)

\bibitem{pham2023svmorph}
Pham, J., Wyetzner, S., Pfaller, M.R., Parker, D.W., James, D.L., Marsden, A.L.: svmorph: Interactive geometry-editing tools for virtual patient-specific vascular anatomies. Journal of Biomechanical Engineering  \textbf{145}(3),  031001 (2023)

\bibitem{poletti2022towardsadigitaltwin}
Poletti, G., Antonini, L., Mandelli, L., Tsompou, P., Karanasiou, G.S., Papafaklis, M.I., Michalis, L.K., Fotiadis, D.I., Petrini, L., Pennati, G.: Towards a digital twin of coronary stenting: A suitable and validated image-based approach for mimicking patient-specific coronary arteries. Electronics  \textbf{11}(3), ~502 (2022)

\bibitem{qiao2022generativeageheart}
Qiao, M., Basaran, B.D., Qiu, H., Wang, S., Guo, Y., Wang, Y., Matthews, P.M., Rueckert, D., Bai, W.: Generative modelling of the ageing heart with cross-sectional imaging and clinical data. In: International Workshop on Statistical Atlases and Computational Models of the Heart. pp. 3--12. Springer (2022)

\bibitem{qiao2023cheart}
Qiao, M., Wang, S., Qiu, H., de~Marvao, A., O'Regan, D.P., Rueckert, D., Bai, W.: Cheart: A conditional spatio-temporal generative model for cardiac anatomy. arXiv preprint arXiv:2301.13098  (2023)

\bibitem{ralapanawa2021epidemiologycoronary}
Ralapanawa, U., Sivakanesan, R.: Epidemiology and the magnitude of coronary artery disease and acute coronary syndrome: a narrative review. Journal of epidemiology and global health  \textbf{11}(2), ~169 (2021)

\bibitem{sato2000teasar}
Sato, M., Bitter, I., Bender, M.A., Kaufman, A.E., Nakajima, M.: Teasar: tree-structure extraction algorithm for accurate and robust skeletons. In: Proceedings the Eighth Pacific Conference on Computer Graphics and Applications. pp. 281--449. IEEE (2000)

\bibitem{sawaya2016contemporarybif}
Sawaya, F.J., Lef{\`e}vre, T., Chevalier, B., Garot, P., Hovasse, T., Morice, M.C., Rab, T., Louvard, Y.: Contemporary approach to coronary bifurcation lesion treatment. JACC: Cardiovascular Interventions  \textbf{9}(18),  1861--1878 (2016)

\bibitem{shit2021cldice}
Shit, S., Paetzold, J.C., Sekuboyina, A., Ezhov, I., Unger, A., Zhylka, A., Pluim, J.P., Bauer, U., Menze, B.H.: cldice-a novel topology-preserving loss function for tubular structure segmentation. In: Proceedings of the IEEE/CVF Conference on Computer Vision and Pattern Recognition. pp. 16560--16569 (2021)

\bibitem{kimimaro2021}
Silversmith, W., Bae, J.A., Li, P.H., Wilson, A.: Kimimaro: Skeletonize densely labeled 3d image segmentations (Sep 2021). \doi{10.5281/zenodo.5539913}, \url{https://doi.org/10.5281/zenodo.5539913}

\bibitem{sonck2022clinical}
Sonck, J., Nagumo, S., Norgaard, B.L., Otake, H., Ko, B., Zhang, J., Mizukami, T., Maeng, M., Andreini, D., Takahashi, Y., et~al.: Clinical validation of a virtual planner for coronary interventions based on coronary ct angiography. Cardiovascular Imaging  \textbf{15}(7),  1242--1255 (2022)

\bibitem{song2023lossguided}
Song, J., Zhang, Q., Yin, H., Mardani, M., Liu, M.Y., Kautz, J., Chen, Y., Vahdat, A.: Loss-guided diffusion models for plug-and-play controllable generation  (2023)

\bibitem{stone2012predictionplaquemorph}
Stone, P.H., Saito, S., Takahashi, S., Makita, Y., Nakamura, S., Kawasaki, T., Takahashi, A., Katsuki, T., Nakamura, S., Namiki, A., et~al.: Prediction of progression of coronary artery disease and clinical outcomes using vascular profiling of endothelial shear stress and arterial plaque characteristics: the prediction study. Circulation  \textbf{126}(2),  172--181 (2012)

\bibitem{straughan2023fully}
Straughan, R., Kadry, K., Parikh, S.A., Edelman, E.R., Nezami, F.R.: Fully automated construction of three-dimensional finite element simulations from optical coherence tomography. Computers in Biology and Medicine  \textbf{165},  107341 (2023)

\bibitem{verhulsdonk2023shapeofmyheart}
Verh{\"u}lsdonk, J., Grandits, T., Costabal, F.S., Krause, R., Auricchio, A., Haase, G., Pezzuto, S., Effland, A.: Shape of my heart: Cardiac models through learned signed distance functions. arXiv preprint arXiv:2308.16568  (2023)

\bibitem{virmani2006pathologycoronaryplaque}
Virmani, R., Burke, A.P., Farb, A., Kolodgie, F.D.: Pathology of the vulnerable plaque. Journal of the American College of Cardiology  \textbf{47}(8S),  C13--C18 (2006)

\bibitem{wang2022semanticdiffusion}
Wang, W., Bao, J., Zhou, W., Chen, D., Chen, D., Yuan, L., Li, H.: Semantic image synthesis via diffusion models. arXiv preprint arXiv:2207.00050  (2022)

\bibitem{wiesner2023generativemodellivingcells}
Wiesner, D., Suk, J., Dummer, S., Ne{\v{c}}asov{\'a}, T., Ulman, V., Svoboda, D., Wolterink, J.M.: Generative modeling of living cells with so (3)-equivariant implicit neural representations. arXiv preprint arXiv:2304.08960  (2023)

\bibitem{zhao2021patientbifurcationstent}
Zhao, S., Wu, W., Samant, S., Khan, B., Kassab, G.S., Watanabe, Y., Murasato, Y., Sharzehee, M., Makadia, J., Zolty, D., et~al.: Patient-specific computational simulation of coronary artery bifurcation stenting. Scientific reports  \textbf{11}(1),  16486 (2021)

\end{thebibliography}

\appendix
\section{Supplementary Materials Overview}
\begin{itemize}
    \item In \cref{appdx:Dataset}, we provide details on dataset curation and augmentation
    \item In \cref{appdx:Topo_loss}, we detail the mathematical formulation of our topological loss.
    \item In \cref{appdx:Features}, we discuss the extraction process for te morphological features and the lumenal skeleton.
    \item In \cref{appdx:Implementation}, we provide implementation details for our autoencoder and diffusion model
    \item In \cref{appdx:Regressors}, we provide implementation details for the neural networks used for morphological and skeletal regression.
    \item In \cref{appdx:Add_Results}, we visualize the effect of adaptive null guidance with each conditioning mechanism in isolation. .
    \item In \cref{appdx:Virtual_Angio}, we provide details on our virtual angioplasty simulations which demonstrated that synthetic segmentation maps can be used for numerical simulation studies.
\end{itemize}

\section{Dataset Processing}
\label{appdx:Dataset}
Our study used 62 CCTA images obtained from the Precise Percutaneous Coronary Intervention Plan (P3) Study \cite{sonck2022clinical}. All data has been anonymized and approved for use by an IRB. The 3D lumen and vessel wall morphology were extracted using a methodology developed by HeartFlow Inc \cite{sonck2022clinical}, and represented as 3D signed distance fields (SDFs) with a resolution of 0.25 mm. To obtain a straightened representation of the coronary artery, a virtual catheter was propagated along the main arterial branch with atherosclerotic lesions for each patient, capturing 2D cross-sections of the lumen and vessel wall at intervals of 0.2 mm. These 2D cross-sections were stacked to construct 3D segmentation maps of the lumen and vessel wall. The total number of 2D frames extracted was 28,533, where the average number of frames per artery was 460. For the segmentation of calcified tissue, thresholding was applied to the corresponding 2D frames extracted by the virtual pullback. The threshold was determined as the greater value between 450 Hounsfield units (HU) or the mean lumen density increased by one standard deviation. The anisotropic resolution of the segmentation maps was set at 60 $\mu$m in-plane and 0.2 mm out-of-plane to capture the longitudinal variation inherent in the arterial structure. To obtain the 3D segmentation maps that are fed into the VAE, we extract 3D patches consisting of 128 frames along the  artery with a random starting frame. We further apply random rotations about the longitudinal axis to augment our dataset.

\section{Topological Loss Implementation}
\label{appdx:Topo_loss}
Our topological loss attempts to penalize topologically incorrect voxels within our autoencoded coronary segmentation maps. We specifically enforce that the vessel wall contains both lumen and calcium within any 2D frame. To enforce such constraints, we use the topological loss as described by Gupta et al. \cite{gupta2022learningtopo} and adapt it to a compound multi-slice context, applying the loss to each 2D cross-section of the coronary segmentation map.
Let $\mathbf{F} \in\mathbb{R}^{C\times H\times W\times D}$ be the multi-class segmentation map predicted by the VAE. To enforce that label-B contains label-A, we take a cross-sectional 2D label map $\textbf{P} \in\mathbb{R}^{H\times W}$ and aim to find a critical pixel map that contains label-A pixels with a label-C neighbor and vice versa, where label-C is specifically defined as the union of all labels that are not label-A or label-B.
First, each label is expanded by 3 pixels using a convolutional kernel $K$ to obtain the neighborhood information $N$.
\begin{equation}
    N_A:=\textbf{P}_A\circledast K \quad\mathrm{and}\quad N_C:=\textbf{P}_C \circledast K,
\end{equation}
where K is set as a 2D 4-connectivity kernel. We then use the neighborhood information to find the critical pixels of each mask ($\textbf{V}_A$ and $\textbf{V}_C$) that fall into the other masks neighborhood.
\begin{equation}
    \textbf{V}_A:=\textbf{P}_A \odot N_C \quad\mathrm{and}\quad \textbf{V}_C:=\textbf{P}_C \odot N_A \quad\mathrm{and}\quad \textbf{V}:=\textbf{V}_A \oplus \textbf{V}_C,
\end{equation}
where $\odot$ represents the Hadamard product and $\oplus$ represents the union operation. 
We set the ground truth label map $\mathbf{G} \in\mathbb{R}^{ H\times W\times D}$ as the vessel wall and use the critical voxel map $\textbf{V}_{3D}$ with a voxel-wise cross-entropy loss $\mathcal{L}_{ce}$
\begin{equation}
    \mathcal{L}_T=\mathcal{L}_{ce}(\mathbf{F}\odot \mathbf{V}_{\text{3D}},\mathbf{G} \odot \mathbf{V}_{\text{3D}}),
\end{equation}
where the critical voxel map $\mathbf{V}_{\text{3D}}$ is found by stacking the critical pixel map $\mathbf{V}$ for each cross section along the depth dimension $D$.

\section{Morpho-Skeletal Feature Extraction}
\label{appdx:Features}
\subsection{Morphological Features}
\label{appdx:Morph_Features}
Morphological features were extracted from multi-channel coronary artery segmentation maps to quantify both 3D and 2D metrics and are detailed in \cref{tab:morph_metrics_eval}. Volumes and cross-sectional areas were calculated from the mean number of positive voxels, while the stenosis ratio was calculated as the minimum lumen area divided by the average lumen volume. Vessel burden was defined as the ratio between vessel and lumen area at the site of minimum stenosis. Plaque was defined as the entire vessel cross-section, consisting of the lumen, vessel wall, and calcium. Circularity was computed as the ratio of area to the square of the perimeter. Thickness was obtained by doubling the maximum value of the Euclidean distance field for the 2D segmentation map. Arclength was calculated as the maximum angular span of the segmentation map contours per cross-sectional area. The centroid was measured as the distance between the centroid and the center of the image. When a morphological feature vector is used for conditioning, the feature curve was smoothed along the length of the artery with a Savitzky-Golay filter using a window length of 21, corresponding to approximately 16\% of the feature vector length. 

\begin{table}[ht]
\centering
\noindent 
\caption{Recorded morphological metrics for evaluation}
\begin{minipage}[t]{0.5\textwidth}
\centering
\begin{tabular}{l}
\toprule
\textbf{2D Morph. Metrics} \\
\midrule
1) Lumen area \\
2) Vessel area \\
3) Plaque area\\
4) Plaque centroid \\
5) Plaque circularity \\
6) Calcium area \\
7) Calcium thickness \\
8) Calcium arclength \\
\bottomrule
\end{tabular}
\end{minipage}%
\begin{minipage}[t]{0.5\textwidth}
\centering
\begin{tabular}{l}
\toprule
\textbf{3D Morph. Metrics} \\
\midrule
1) Lumen volume \\
2) Stenosis ratio \\
3) Vessel volume \\
4) Vessel burden \\
5) Minimum plaque circularity \\
6) Mean plaque circularity \\
7) Calcium volume \\
8) Calcium length \\
9) Maximum calcium thickness \\
10) Mean calcium thickness \\
11) Maximum calcium arclength \\
12) Mean calcium arclength \\
\bottomrule
\end{tabular}
\end{minipage}

\label{tab:morph_metrics_eval}
\end{table}

\subsection{Skeletal Features} 

\begin{table}[h]
\centering
\begin{minipage}{0.5\textwidth}
\centering
\caption{Hard Skel. Parameters}
\begin{tabular}{c|c}
\hline
\textbf{Parameter} & \textbf{Value} \\ \hline
Const. & 10 \\
Scale & 1.5 \\
Prdf scale & 1e5 \\
Prdf exponent & 5 \\
Tick threshold & 10 \\
\label{tab:hard_skel_config}
\end{tabular}
\end{minipage}%
\begin{minipage}{0.5\textwidth}
\centering
\caption{Skel. Processing Parameters}
\begin{tabular}{c|c}
\hline
\textbf{Parameter} & \textbf{Value} \\ \hline
$\sigma_{\text{skel}}$ & 1 \\
Gaussian Kernel Size & 3 \\
Max pool Kernel Size & 4 \\
Max pool Stride & 4 \\
\label{tab:skel_proc_config}
\end{tabular}
\end{minipage}
\end{table}

\label{appdx:Skel_Features}
Our skeletal regressors consist of 1) a skeletonization step, which takes in a lumenal segmentation map, and 2) a pre-processing step to enable concatenation with the encoded latent representation. We leverage three types of skeletonization: non-differentiable (hard), differentiable (soft), and neural skeletonization. Our non-differntiable skeletonization process uses the Kimimaro library \cite{kimimaro2021}, which employs the TEASAR algorithm \cite{sato2000teasar}, to derive a high-quality skeleton. Our differentiable skeletonization process leverages an iterative soft skeletonization algorithm by Shit et al. \cite{shit2021cldice}, which is applied on lumen segmentation maps. Our neural skeletonization is done by a neural network trained to regress the processed output of the non-differentiable skeletal regressor (see \cref{appdx:Regressors}).

For hard skeletonization, the pre-processing step includes conversion into a binary voxel grid, Gaussian blurring with a standard deviation denoted by $\sigma_{\text{skel}}$, and normalization. To reduce the grid size, we employ max pooling and subsequently resize the skeleton map to fit the dimensions of the encoded latent representation. For soft skeletonization, we downsample the input lumen segmentation map to a size of $H/2 \times W/2 \times D$ before skeletonization to conserve GPU memory. The resulting skeleton is then upscaled to its original resolution and is then processed similarly to hard skeletonization. For neural skeletonization, the processing step is equivalent to that of hard skeletonization. The configuration details for the hard skeletonization and associated processing can be found in \cref{tab:hard_skel_config,tab:skel_proc_config} respectively.

The outputs of our skeletonization methods along with their processed outputs are shown in \cref{fig:skel_val}. Our hard skeletonization method creates intact skeletons given an input lumen segmentation map. Processing the skeletons with Gaussian blurring and max pooling retains the skeletal connectivity after resizing to a coarser resolution. In contrast, the soft skeletonization method produces disconnected skeletal branches, which may or may not be repaired after processing.

\section{Latent Diffusion Model Implementation}
\label{appdx:Implementation}
We adapt the VAE and LDM architectures specified by Kadry et al. \cite{kadry2023probing} to the case of coronary segmentation map generation. The input and output channels of the VAE are set to a value of 4 to correspond to the number of tissues. The number of input channels for the LDM was set to a value of 6 such that 3 channels correspond to the encoded latent and 3 channels correspond to two morphological conditioning maps and one skeletal conditioning map. The hyperparameters and training configuration for the VAE and LDM are contained in \cref{tab:VAE_config} and \cref{tab:LDM_config} respectively. 
\begin{table}[h]
\centering
\begin{minipage}{0.5\textwidth}
\centering
\caption{VAE Config.}
\begin{tabular}{c|c}
\hline
\textbf{Hyperparameter} & \textbf{Value} \\ \hline
lr &\hspace{1pt} $1 \times 10^{-5}$ \\
Iterations &$4 \times 10^{4}$ \\
Batch Size & 2 \\
Num. Channels & [64, 128, 192] \\
Num. Res. Blocks & 2 \\
Downscaling Factor & 4 \\
$\lambda_{recon}$ & 1 \\
$\lambda_{KL}$ & $1 \times 10^{-6}$ \\
$\lambda_{topo}$ & $2 \times 10^{-4}$ \\
\end{tabular}
\label{tab:VAE_config}
\end{minipage}%
\begin{minipage}{0.5\textwidth}
\centering
\caption{LDM Config.}
\begin{tabular}{c|c}
\hline
\textbf{Hyperparameter} & \textbf{Value} \\ \hline
\textbf{Training} & \\ \hline
lr & $1 \times 10^{-5}$ \\
Iterations &  $5 \times 10^{3}$ \\
Batch Size & 1 \\
Num. Channels & [128, 256, 384] \\
Num. Res. Blocks & 2 \\
Num. Attn. Heads & 1 \\
Attn. Res. & 8, 4, 2 \\
$\sigma_{\text{data}}$ & 1 \\
$p(\sigma)$ mean & 1 \\
$p(\sigma)$ std & 1.2 \\
\hline
\textbf{Sampling} & \\
\hline
$\sigma_{min}$ & $1 \times 10^{-2}$ \\
$\sigma_{max}$ & 80 \\
$\rho$ & 3 \\
\end{tabular}
\label{tab:LDM_config}
\end{minipage}
\end{table}

\section{Morpho-skeletal Regressor Implementation}
\label{appdx:Regressors}
\begin{figure*}[h]
  \centering
  \includegraphics[width=0.8\textwidth]{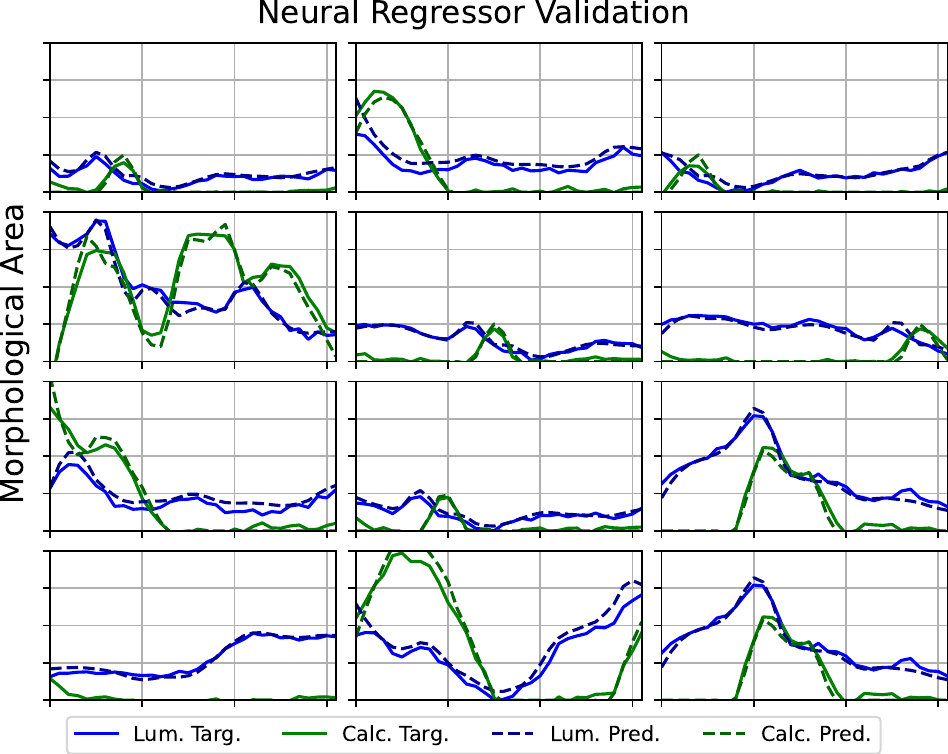}
  \caption{Comparison of target morphological areas against predictions by our neural morphological regressor. Our regressor can predict the morphological features associated with coronary segmentation maps to a high degree of fidelity.}
  \label{fig:morph_val}
\end{figure*}

\subsection{Morphological Regressor Training} The neural morphological regressor $\mathcal{\bar{M}}_\phi$ consists of a 3D encoder, an adaptive pooling layer, and an MLP head. To regress the morphological features, we first encode the coronary segmentation map $\mathbf{x}$ into a latent representation of size $\alpha \times h \times w \times d$ where $\alpha$ is the number of latent channels. This latent representation is then processed through an adaptive average pooling layer, resulting in a feature map of dimensions $\alpha \times 1 \times 1 \times d$. Subsequently, an MLP head with two fully connected layers is applied to transform this feature map into 1D morphological features of size $m \times d$, where $m$ is the number of regressed morphological features. The output is then resized to match the output shape of our original morphological regressor $\mathcal{M}$. \cref{tab:morph_regressor_config} details the hyperparameters and training details for the morphological regressor. The network is trained to minimize a morphological regression loss $\mathcal{L}_{\text{morph}}$ with the Adam optimizer.
\begin{equation}
    \mathcal{L}_{\text{morph}}=\mathcal{L}_{\text{MSE}}(\mathcal{M}(\mathbf{x}),\mathcal{\bar{M}_\phi(\mathbf{x}))}
\end{equation}
The performance of our trained morphological regressor can be seen in \cref{fig:morph_val}, in which a regressor is applied to coronary segmentation maps drawn from the validation set. The predictions from our morphological regressor show good correlation with the ground truth morphology as measured by the ground truth morphological regressor $\mathcal{M}$.

\subsection{Skeletal Regressor Training} 
\begin{figure*}[h]
  \centering
  \includegraphics[width=\textwidth]{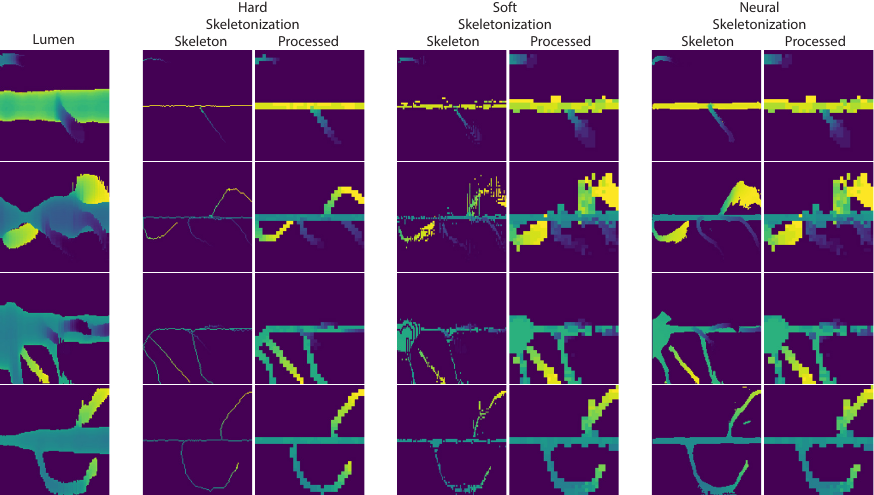}
  \caption{Comparison of different skeletonization methods before and after processing. Left column displays lumen depth maps from validation set. Center left, center right, and rightmost columns indicate skeletonization outputs for hard, soft, and neural regressors respectively. Hard skeletonization outperforms both soft skeletonization and the neural skeletal regressor.}
  \label{fig:skel_val}
\end{figure*}

For the skeletal regressor  $\mathcal{\bar{S}}_\psi$, we employ the SegResNet architecture \cite{myronenko20193dsegresnet}, a U-net based architecture designed to regress the skeletal heatmap for an input coronary segmentation map $\mathbf{x}$. Following the decoder stage, the resulting skeleton voxel grid undergoes blurring and maxpooling similar to the output of the hard skeletonization process. \cref{tab:skel_regressor_config} details the hyperparameters and training details for the skeletal regressor. The network is trained to minimize a skeletal regression loss $\mathcal{L}_{\text{skel}}$ with the Adam optimizer.
\begin{equation}
    \mathcal{L}_{\text{skel}}=\mathcal{L}_{\text{MSE}}(\mathcal{S}(\mathbf{x}),\mathcal{\bar{S}_\psi(\mathbf{x}))}
\end{equation}
The skeletal regressor predictions are visualized in the rightmost column of \cref{fig:skel_val}. It can be seen that while the regressor was trained on the processed output of the hard skeletonization method after pre-processing, it has learned to regress skeletons that are similar to the soft skeletonization process, but with a higher degree of continuity.

\begin{table}[h]
\centering
\begin{minipage}{0.5\textwidth}
\centering
\caption{Morph. Regressor Config.}
\begin{tabular}{c|c}
\hline
\textbf{Parameter} & \textbf{Value} \\ \hline
lr & $1 \times 10^{-4}$ \\
Iterations & 4000 \\
Batch size & 1 \\
Num. Channels & [32, 64, 96] \\
Num. Res. Blocks & 2 \\
Num. Latent Channels & 64 \\
MLP Channels & 1024 \\
\label{tab:morph_regressor_config}
\end{tabular}

\end{minipage}%
\begin{minipage}{0.5\textwidth}
\centering
\caption{Skel. Regressor Config}
\begin{tabular}{c|c}
\hline
\textbf{Parameter} & \textbf{Value} \\ \hline
lr & $1 \times 10^{-4}$ \\
Iterations & 4000 \\
Batch size & 1 \\
Encoder Channels & [16, 32, 32, 64] \\
Decoder Channels & [64, 64, 64, 64] \\
\label{tab:skel_regressor_config}
\end{tabular}
\end{minipage}
\end{table}

\section{Additional Ablation Results}
\label{appdx:Add_Results}
In this section, we further study the effect of ablating one or more conditioning signals during LDM training. We use four LDMs corresponding to 1) no-conditioning, 2) morphological conditioning only, 3) skeletal conditioning only, and 4) joint conditioning. For each LDM, we sample 100 segmentation maps with 25 sampling steps, sweeping over a range of guidance weights for all but the unconditional model. Unless stated otherwise, we used adaptive null guidance with a weight of 5, and the target morpho-skeletons were derived from the validation set. In addition to recording metrics for sampling quality and conditional fidelity (\cref{fig:Guidance_Ablation}),  we visualize the distribution of morphological features as a set of 1D KDE plots (\cref{fig:kde_dng,fig:kde_uncond,fig:kde_dng_skel,fig:kde_dng_morph}) and visualize the segmentation maps in \cref{fig:Add_labelmaps}.
\subsubsection{Unconditional Generation}
\cref{fig:kde_uncond} shows that unconditional generation can recapitulate the distribution in various morphological features in the training set, with good coverage compared to the validation set. Unconditionally generated segmentation maps can be seen in \cref{fig:Add_labelmaps}, where unconditional sampling can synthesize a wide array of morphological and skeletal features.

\subsubsection{Joint Conditioning}
\begin{figure*}[h]
  \centering
  \begin{minipage}[t]{0.45\textwidth} 
    \includegraphics[width=\textwidth]{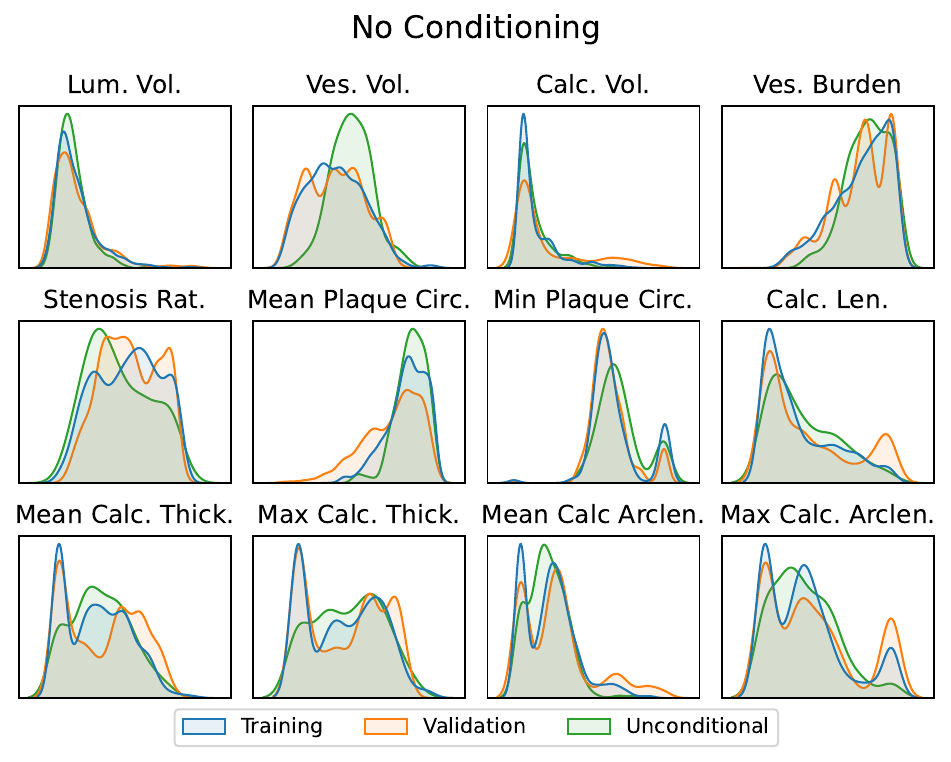}
    \caption{KDE plots of morphological features comparing the training set, validation set, and the set of segmentation maps sampled through unconditional generation.}
    \label{fig:kde_uncond}
  \end{minipage}
  \hfill 
  \begin{minipage}[t]{0.45\textwidth} 
    \includegraphics[width=\textwidth]{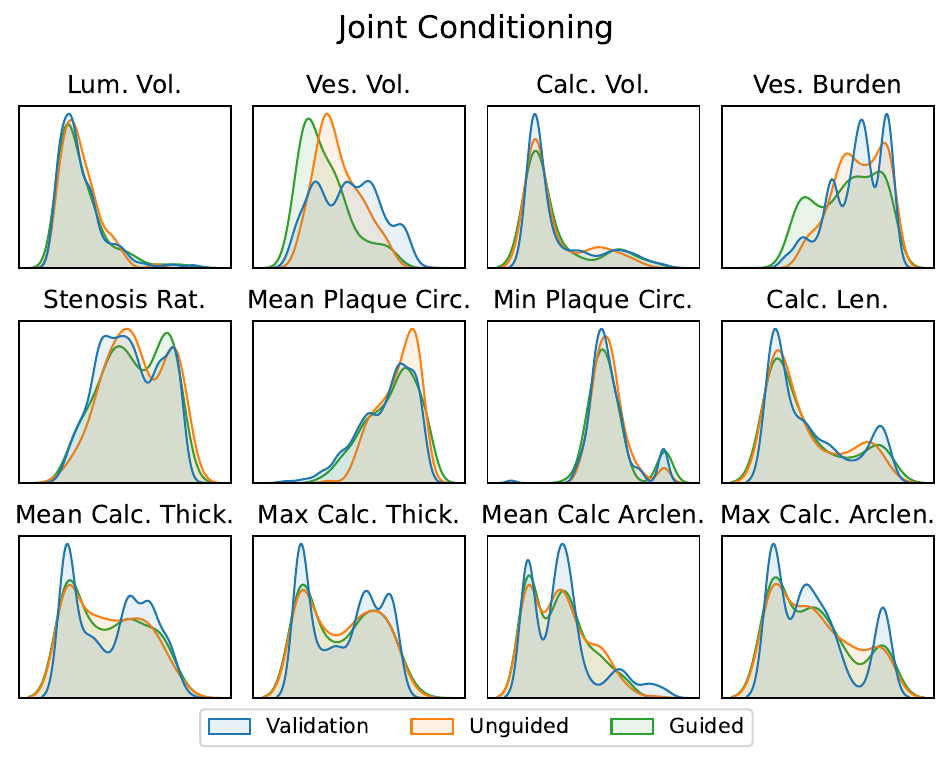}
    \caption{KDE plots comparing the morphological feature distributions of synthetic segmentation maps produced by joint conditioning.}
    \label{fig:kde_dng}
  \end{minipage}
\end{figure*}

\begin{figure*}[h]
  \centering
  \includegraphics[width=0.7\textwidth]{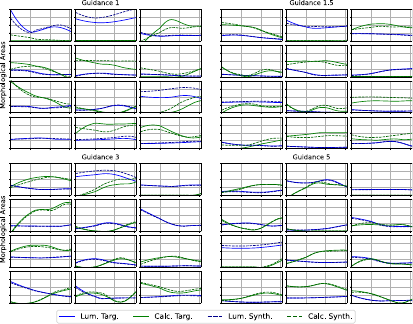}
  \caption{Comparison of target morphological features for lumen and calcium against the equivalent morphological features derived from synthetic segmentation maps using adaptive null guidance at various guidance weights. Our guidance method enhances morphological conditional fidelity as guidance weight increases.}
  \label{fig:skel_synth}
\end{figure*}

\begin{figure*}[h]
  \centering
  \includegraphics[width=0.7\textwidth]{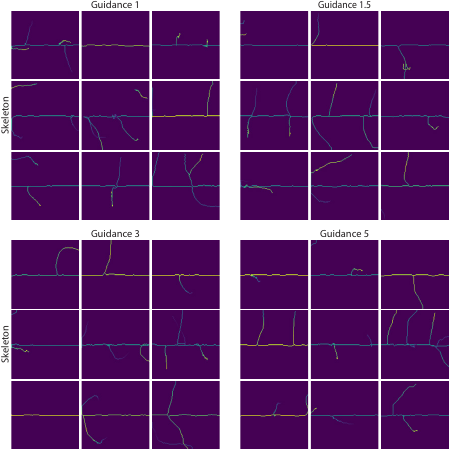}
  \caption{Comparison of skeleton depth maps derived from synthetic segmentation maps using adaptive null guidance at various guidance weights.}
  \label{fig:morph_synth}
\end{figure*}

\cref{fig:kde_dng} shows that guidance improves the similarity of the synthetic distribution to the validation distribution for most morphological features. This can also be seen in \cref{fig:morph_synth}, where guidance increases morphological conditioning fidelity for a variety of target morphological features. Similarly, \cref{fig:skel_synth} displays skeleton depth maps for various adaptive null guidance values. Our joint conditioning framework can produce a wide variety of skeletal configurations without components that are disconnected from the main branch. Synthetic segmentation maps produced by joint conditioning can be seen in \cref{fig:Add_labelmaps}, which exhibits a wide array of morphological and skeletal features.

\begin{figure*}[h]
  \centering
  \begin{minipage}[t]{0.49\textwidth}
    \includegraphics[width=\textwidth]{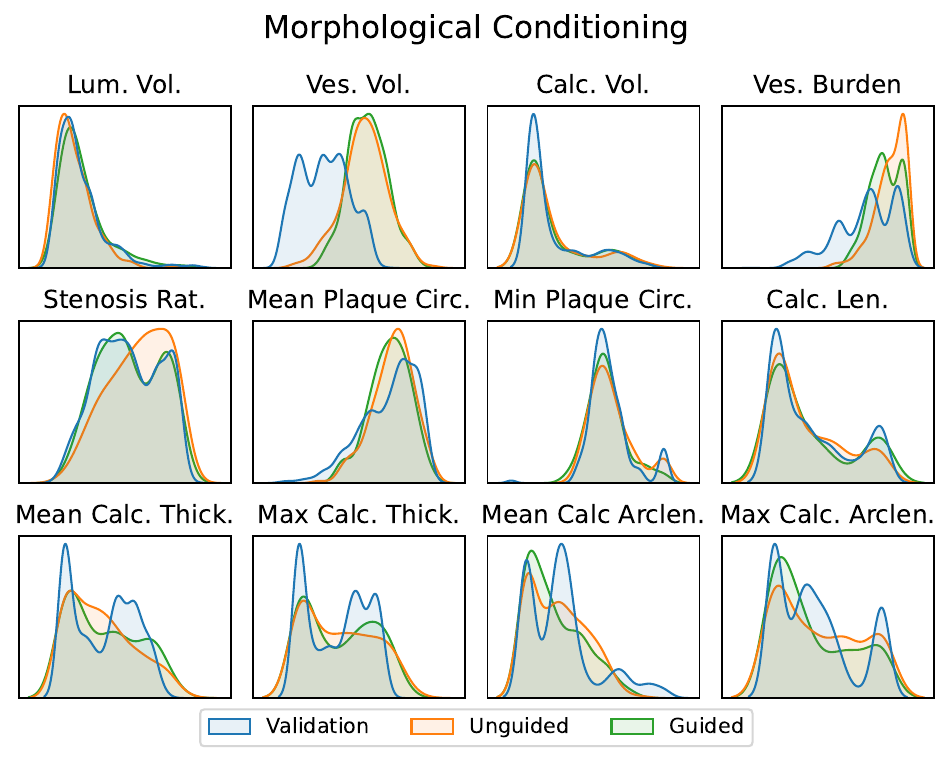}
    \caption{KDE plots comparing the morphological feature distributions of synthetic segmentation maps produced by morphological conditioning alone.}
    \label{fig:kde_dng_morph}
  \end{minipage}
  \hfill
  \begin{minipage}[t]{0.49\textwidth}
    \includegraphics[width=\textwidth]{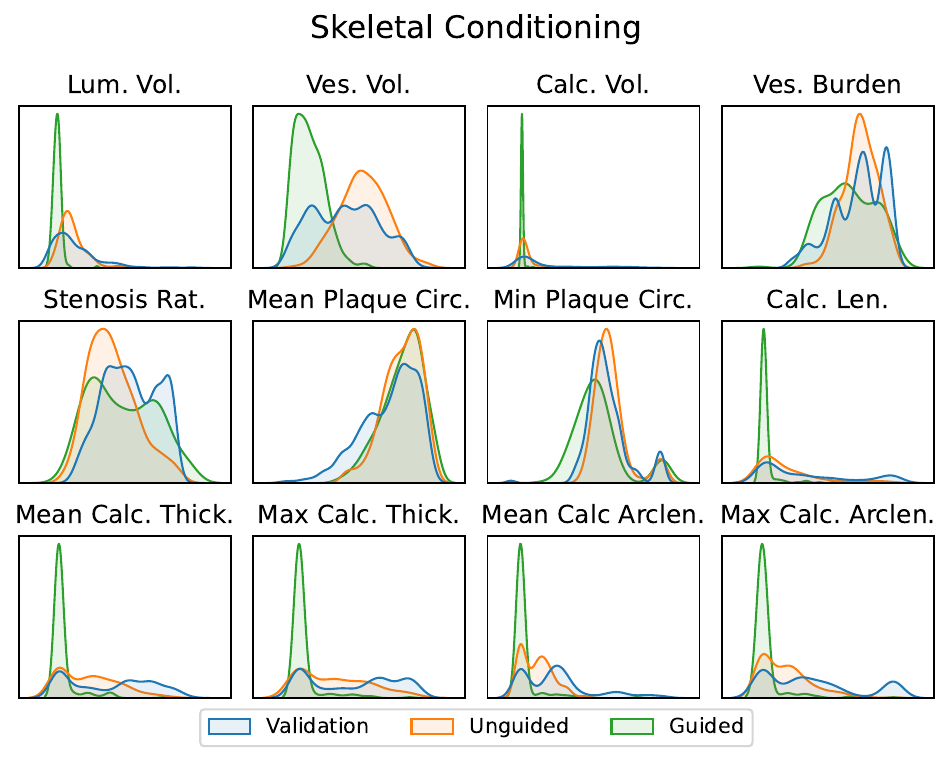}
    \caption{KDE plots comparing the morphological feature distributions of synthetic segmentation maps produced by skeletal conditioning alone.}
    \label{fig:kde_dng_skel}
  \end{minipage}
\end{figure*}

\subsubsection{Morphological Conditioning Only}
\cref{fig:kde_dng_morph} shows that morphological guidance slightly improves the similarity of between the morphological distributions of synthetic and validation segmentation maps. This can be seen in \cref{fig:Guidance_Ablation}, which shows enhanced morphological conditioning fidelity with increasing guidance strength, similar to joint guidance. segmentation maps sampled by morphological conditioning can be seen in \cref{fig:Add_labelmaps}, which exhibit non-physiological bifurcation structures.

\subsubsection{Skeletal Conditioning Only}
\begin{figure*}[h]
  \centering
  \includegraphics[width=\textwidth]{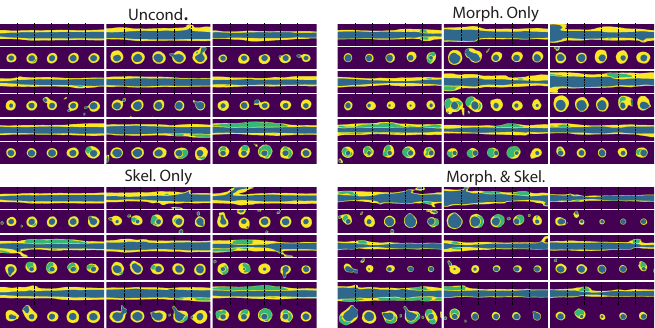}
  \caption{Example segmentation map cross sections for unconditional sampling, morphological conditioning only, skeletal conditioning only, and joint morphological and skeletal conditioning. No guidance was used during sampling.}
  \label{fig:Add_labelmaps}
\end{figure*}

\begin{figure}[h]
  \centering
  \includegraphics[width=\linewidth]{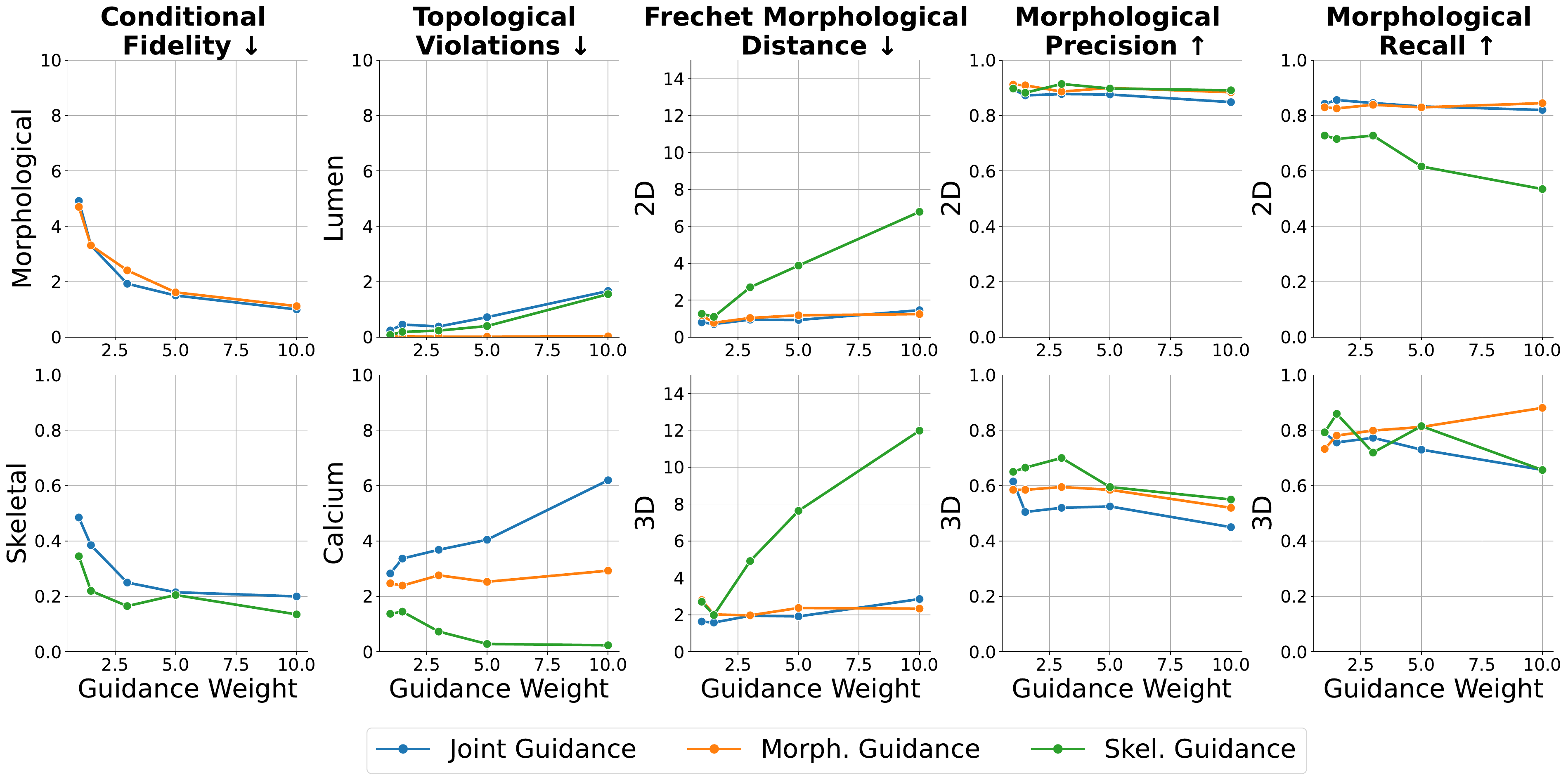}
  \caption{Conditional fidelity and sampling quality metrics for adaptive null guidance using morphological, skeletal, and joint conditioning.}
  \label{fig:Guidance_Ablation}
\end{figure}

\cref{fig:kde_dng_skel} shows that unguided skeletal conditioning does not significantly change the morphological distribution of features as compared to the validation set. However, with skeletal guidance, the morphological distributions shift towards exhibiting smaller lumen, vessel, and calcium volumes. \cref{fig:Guidance_Ablation} shows that skeletal guidance alone can slightly improve skeletal fidelity as compared to joint guidance. This comes at the cost of reduced recall and increased Frechet distance, likely due to the morphological bias towards smaller volumes. Segmentation maps produced by skeletal conditioning can be seen in \cref{fig:Add_labelmaps}, where skeletal conditioning produces a variety of realistic bifurcation structures.

\label{appdx:Ablations}

\section{Virtual Angioplasty Simulation}
\label{appdx:Virtual_Angio}
To demonstrate the compatibility of our synthetic arteries with numerical simulation, we simulate a virtual intervention one two arteries generated from a patient-specific morpho-skeleton. Our synthetic segmentation map is processed by the methodology outlines by Straughn et al. \cite{straughan2023fully}, in which the lumen, vessel wall, and calcium labels are morphologically interpolated and processed to produce an isotropic segmentation map which is used to produce a 3-dimensional multi-material tetrahedral mesh. We then utilize the 3D mesh for virtual angioplasty simulations using the finite element method through the commercial code ABAQUS Explicit \cite{abaqus2023}. Briefly, the simulation platform consists of the coronary artery, stent, and deployment balloon. The artery materials included vessel wall, which was defined as a relatively soft hyperelastic-plastic material, and calcium, which was defined as a stiff elastic material \cite{straughan2023fully}. The stent was geometrically modelled with hexahedral elements and exhibits an elasto-plastic constitutive law \cite{poletti2022towardsadigitaltwin}. The balloon was geometrically modelled with shell elements and exhibits a hyperelastic constitutive law \cite{poletti2022towardsadigitaltwin}. The first stage of the simulation consists of stent crimping, in which the stent begins at a nominal diameter of 3mm and is compressed to a smaller diameter of 1mm with rigid crimping planes. In the second stage, the stent and a folded balloon are inserted into the straightened artery model, at which point the balloon is inflated to a pressure of 14 atm, expanding the stent into the arterial wall.

\end{document}